\documentclass{aa}  

\usepackage{graphicx}
\usepackage{txfonts}
\usepackage{natbib}
\usepackage{xcolor}  
\usepackage{makecell}

\usepackage{hyperref}
\begin{document}

   \title{Intergalactic Magnetic Field constraints from detected very high-energy Gamma-Ray Bursts using the Cherenkov Telescope Array Observatory}

   \author{T. Keita
          \inst{1}
          \and
          R. Belmont\inst{2}
          \and
          Th. Stolarczyk\inst{1}
          }

   \institute{Université Paris-Saclay, Université Paris Cité, CEA, CNRS, AIM, 91191, Gif-sur-Yvette, France\\
              \email{teneman.keita@cea.fr}
         \and
             Université Paris Cité, Université Paris-Saclay, CEA, CNRS, AIM, 91191, Gif-sur-Yvette, France\\
             \email{renaud.belmont@cea.fr}
             }

   \date{Received XXX XX, 20XX; accepted XXX XX, 20XX}

  \abstract
   {Defined as the magnetic field permeating cosmic voids, the Intergalactic Magnetic Field (IGMF) is thought to be a relic of the Big Bang, tracing a primordial magnetic seed at the origin of all astrophysical fields. Yet, it has thus far escaped detection. Lower limits on the IGMF strength can be established by observing very high-energy (VHE) photons from extragalactic sources. Specifically, this can be achieved by characterising the time-delayed secondary emission induced by highly energetic transient sources, such as gamma-ray bursts (GRBs). Most studies exclude values of the IGMF below $10^{-17}\;\mathrm{G}$ by comparing the expected effect to the sensitivity curves of various instruments in the $\mathrm{GeV}$ range or above. In this work, we simulate CTAO observation data under realistic observation conditions and perform spectral-temporal fits to estimate the constraints CTAO will bring on the IGMF once fully deployed. We apply the methodology to simulated sources with properties comparable to the few GRBs detected at VHE. In particular, we show that CTAO will probe strengths up to $\sim 10^{-15}\;\mathrm{G}$ when detecting sources similar to GRBs 221009A and 190114C. We also show that existing observations of GRB 221009A by the first CTAO Large Sized Telescope LST-1 favour a strength of $3\times 10^{-17}\;\mathrm{G}$.}
 

   \titlerunning{Constraints on the IGMF with GRB afterglows and CTAO} 
   \maketitle

\section{Introduction}
\label{Intro}
\quad Magnetic fields have been detected at every observable scale, in stars (\citealp{Donati_2009}), galaxies (\citealp{Bernet_2008}), galaxy clusters (\citealp{Feretti2012}) and galaxy filaments (\citealp{Kronberg2006}). Producing such astrophysical magnetic fields from non-magnetised media is a challenge to modern physics. However, they could naturally result from the amplification of a primordial, weak magnetic field through compression by converging flows and/or dynamo processes. In contrast to strong astrophysical magnetic fields, such a primordial field could be generated during the inflation phase  (\citealp{Tripathy_2023} and \citealp{Cecchini_2023}) by the coupling of the inflaton field with electromagnetism, or during electroweak (\citealp{Joyce_1997} and \citealp{Huber_2006}) and QCD phase transitions (\citealp{Schwarz:2009ii} and \citealp{Boeckel_2012}). Within cosmic voids, this primordial field evolves under the combined effects of cosmic expansion and magnetic dissipation, leading to the formation of the Intergalactic Magnetic Field (IGMF). The IGMF is often described as a turbulent field with a mean strength \( B \) ranging from $10^{-17}$ to $10^{-9}\;\mathrm{G}$, and an average perturbation scale (a.k.a. correlation length) \( \lambda_B \)~(see \citealp{Durrer_2013}). Due to theoretical predictions based on the early magnetic dissipation theory (\citealp{banerjee_2002}), this correlation length is expected to range from the sub-parsec scale to megaparsecs depending on the field strength.
    
If contamination from the magnetised outflows of galaxies and galaxy clusters is negligible (\citealp{Blunier_2024}), constraints on the IGMF can be used to infer the physics of the primordial universe. In particular, it may serve as indirect evidence for inflation (\citealp{Tripathy_2023}) or first-order phase transitions. The latter would, in turn, constrain the leptogenesis models and the matter-antimatter asymmetry (\citealp{Joyce_1997}). By analysing the Cosmic Microwave Background (CMB) temperature and polarisation anisotropies, the Planck collaboration established a conservative upper limit of $B < 5.6 \times 10^{-9}\;\mathrm{G}$ (\citealp{Ade_2016}). Combining these CMB anisotropy data and MHD simulations, \cite{Jedamzik_2019} derived the most stringent upper limits, at $4.7\times 10^{-11}\;\mathrm{G}$. More recently, the NANOGrav collaboration announced the discovery of a Stochastic Gravitational Wave Background (\citealp{Agazie_2023}). This signal is compatible with a QCD phase transition scenario (\citealp{Roper_Pol_2022}) inducing a primordial IGMF with strength between $5\times10^{-12}$ and $10^{-11}\;\mathrm{G}$ (\citealp{Jedamzik_2025}). Furthermore, the presence of a magnetic field of this magnitude would increase baryon inhomogeneities before recombination, shifting the estimated value of the Hubble constant derived from CMB data upward. This offers a compelling mechanism to alleviate the Hubble tension (\citealp{Jedamzik_2020}).

Currently, the only way to probe the lowest values of the IGMF strength relies on observations of very-high-energy (VHE) sources like Active Galactic Nuclei (AGN) and Gamma-Ray Bursts (GRBs) and the search for a delayed secondary emission at low energies. Indeed, VHE gamma rays interact after a few hundreds $\mathrm{Mpc}$ with the Extragalactic Background Light (EBL), leading to the so-called EBL absorption. This absorption creates an exponential cut-off at few hundred of $\mathrm{GeV}$ in the source spectrum.  In this interaction, a relativistic electron-positron pair is produced. Both leptons upscatter CMB photons, typically up to hundreds of $\mathrm{GeV}$ or $\mathrm{TeV}$ depending on the original gamma-ray energy. They can be absorbed again, creating new pairs, hence new photons and so on, generating an electromagnetic cascade. Provided the primary photons energies do not exceed $100\;\mathrm{TeV}$, the first generation of secondary photons is not significantly absorbed and the cascade is well approximated by a single generation description as sketched in Fig. \ref{Triangle}. The typical cascade signature is a spectral excess searched at lower energy. As the charged pairs are deflected by the IGMF, the secondary photons detected on Earth are produced off the line of sight and observed in a \textit{pair halo} (\citealp{Neronov_2007}) while arriving with a time delay, a \textit{pair echo} (\citealp{1995Nature}). So far, searches for halos or echoes have mainly resulted in upper limits on the secondary emission flux. Because the IGMF spreads the emission in space and time, these observational constraints can be interpreted as lower bounds on the magnetic field strength. The energy range suited to IGMF studies is currently covered by Fermi-LAT (\citealp{Ackermann_2012}), the three present Imaging Atmospheric Cherenkov Telescopes (IACTs), H.E.S.S. (\citealp{Puhlhofer_2023}), VERITAS (\citealp{Hanna_2024}), MAGIC (\citealp{Blanch_2024}) and, more recently, the LHAASO (\citealp{Cao_2022}) detector.

AGNs are relatively persistent sources that can emit continuously for tens to millions of years. These sources are usually used to characterise the pair halo. The absence of pair halos in blazars is the main method for establishing lower limits on the IGMF strength. Early searches by the H.E.S.S. collaboration (\citealp{Abramowski_2014}) pioneered this technique, excluding strengths from $3\times 10^{-16}\;\mathrm{G}$ to $10^{-14}\;\mathrm{G}$ assuming $\lambda_B \gtrsim 1\;\mathrm{Mpc}$. Later, the non-detection of any halo from the hard-spectrum blazar 1ES 1218+304 allowed the VERITAS collaboration (\citealp{Archambault_2017}) to exclude a range of IGMF strengths between $5 \times 10^{-15}\;\mathrm{G}$ and $7 \times 10^{-14}\;\mathrm{G}$. Complementary constraints have also been derived using pure spectrum information. Notably, \cite{Neronov_2010} reported a lower limit of $B \geq 3 \times 10^{-16}\;\mathrm{G}$ based on the non-detection of secondary emission in the Fermi-LAT spectra of several blazars. 

However, these limits usually depend on the source activity duration and the jet opening angle, which are both unknown. Furthermore, since the detection relies on a cascade spatial extension, the lower limits are also sensitive to the uncertainties on the instrument Point-Spread Functions (PSF). Last, some authors have argued that plasma instabilities generated by the pair beam could also quench the secondary emission by efficiently cooling the leptons before they can emit secondary photons (see \citealp{Broderick_2012}). More recently, \cite{Acciari_2023} and \cite{Blunier_2026} established weaker but significantly more robust limits, up to $2\times 10^{-17}\;\mathrm{G}$, by discarding the assumption of long-term constant emission, strictly relying instead on the measured source variability and finite observation time.

Since the detection of GRB 190114C by MAGIC (\citealp{2019}), we know that long GRBs can emit $\mathrm{TeV}$ gamma-rays in their afterglow phase, that results from the collision of the relativistic jets of expelled matter with the surrounding interstellar medium (\citealp{foffano2025}). GRB emits very high-energy gamma-rays on time scales that are too short to produce a detectable halo, however, they are ideal targets for pair-echo searches. Most of the assumptions required to use AGNs and pair halos are not required for studying time delays with GRBs: the short time scale constrains the secondary gamma-rays to propagate at an angle well within the PSF of any instrument ($\sim10^{-5}$ degree), and are produced by primaries emitted well within the jet of any GRB. The activity duration of the source is well constrained and much shorter than the time required for hypothetical plasma instabilities to develop (\citealp{Broderick_2012}). In addition to GRB 190114C, four other VHE GRBs have been detected: GRB 180720B (\citealp{Abdalla_2019}), GRB 190829A (\citealp{Hess_2021}), GRB 201216C (\citealp{Abe_2023}) and GRB 221009A (\citealp{Lhaaso_2023}). In both \cite{Dzhatdoev_2020} and \cite{Vovk_2023a} they derived constraints using GRB 190114C,  while \cite{Dzhatdoev_2023}, \cite{Huang_2023} and \cite{Vovk_2023b} analysed GRB 221009A. To date, the best lower limits on the IGMF strength using GRBs and time delays are around $10^{-18}\;\mathrm{G}$, using LHAASO and Fermi-LAT data. 

The Cherenkov Telescope Array Observatory (CTAO) is the next generation of VHE telescopes (\citealp{CTAO_2018}, \url{https://www.ctao.org/}). It is composed of two arrays under construction on the La Palma island (Canary Islands, Spain) and in the Atacama Desert (Paranal, Chile). In the Alpha configuration, the arrays comprise $4$ Large Size Telescopes (LSTs) and $9$ Medium Size Telescopes (MSTs) in the North, and $14$ MSTs and $37$ Small Size Telescopes (SSTs) in the South. LSTs are specially designed for GRB detection with a maximal required slewing time of $30\mathrm{s}$ (MSTs and SSTs are required to slew in less than $90$ and $60\mathrm{s}$ respectively) and are suited to energies between $20\;\mathrm{GeV}$ and $1\;\mathrm{TeV}$. The first LST, LST-1, is already taking data. MSTs are more efficient for the $\mathrm{TeV}$ range ($100\;\mathrm{GeV}$ to $50\;\mathrm{TeV}$), with a sensitivity improvement of a factor of $10$ compared to the former IACT generation. In the South, SSTs cover energies from $1$ to $300\;\mathrm{TeV}$. The angular resolution spans from a fraction of a degree at low energies to $1$ arcmin at $300\;\mathrm{TeV}$, and the field of view diameter from $5$ to $10$ degrees, respectively. Recent studies by \cite{Miceli_2024} and \cite{Xia_2024} compared the expected secondary fluxes for GRB 190114C and GRB 221009A to the published CTAO sensitivity curves and deduced that a lower limit of $10^{-17}\;\mathrm{G}$ is reachable. However, the influence of observational conditions has yet to be considered; moreover, previous studies generally evaluate the primary and cascade emissions separately, failing to address the possible degeneracy between them.

In this article, we aim at quantitative constraints by combining the secondary emission temporal and spectral properties in a single analysis, while accounting for night durations and the evolution of the instrument energy threshold and response with altitude. After briefly introducing the phenomenology of cosmological electromagnetic cascades, we describe the Monte Carlo simulation code used to produce synthetic data from the total emission of a source similar to GRB 190114C and analyse it assuming realistic CTAO observational constraints. We then replicate this work on GRBs 221009A, 190829A and 180720B \footnote{The secondary emission of GRB 201216C is too faint to be detectable under any IGMF assumptions, because of a too high redshift and a very soft energy spectrum that limits the high-energy component, as well as a very low primary flux decay that masks any secondary emission. It will not be analysed here.}. Regarding GRB 221009A, existing data from LST-1 (\citealp{Abe:2024kw}, \citealp{lst1}) are also examined.  Finally, we discuss the implications of our analysis on the current and CTAO-expected IGMF constraints.

\section{Phenomenology and simulations}

This section reviews the physical processes underlying cosmological cascade formation and propagation in the presence of the IGMF, the phenomenological implications for gamma-ray observations, and the numerical methods used to model these effects. We first present a detailed description of the cascade development and show how it leads to observable signatures such as time delays and spectral features. This is followed by a presentation of the simulation framework used to model these processes, and the post-processing techniques applied to extract predictions for various observational scenarios. Furthermore, we describe the generation of synthetic datasets that mimic the expected CTAO response under realistic observing conditions. Finally, we detail the fit methodology used to assess the ability of CTAO to recover IGMF strengths from these simulated observations.

\subsection{Cascade phenomenology}
\label{Theory}

\begin{figure}
    \centering
    \includegraphics[width=1\linewidth]{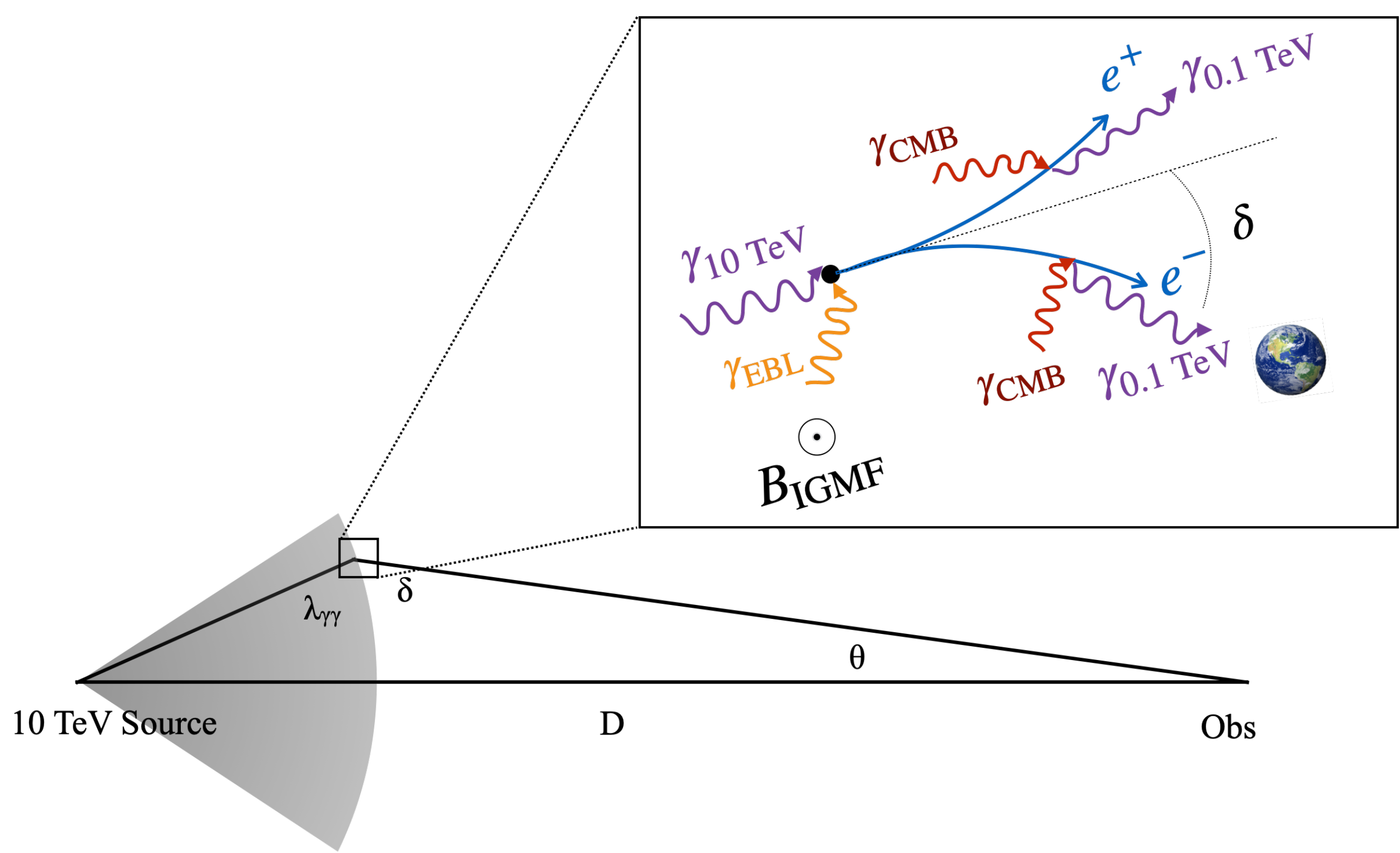}
    \caption{Schematic view of the cascade in the one-generation approximation. Primary photons produce pairs at a distance $\lambda_{\gamma\gamma}$ from the source. The IGMF deflects these pairs by an angle $\delta$, and creates secondary photons observed at an angle $\theta$ with respect to the line of sight. Adapted from \citealp{Keita_2025}.}
    \label{Triangle}
\end{figure}

VHE gamma rays interact with the EBL photons with an absorption length, $\lambda_{\gamma\gamma}$, of the order of $100\;\mathrm{Mpc}$, and produce electron-positron pairs. These relativistic leptons up-scatter CMB photons to high or very high energies with a mean energy proportional to the square of the lepton energy $E_e$:
\begin{equation}
\label{compton}
    \langle E_\gamma \rangle = \frac{4\epsilon_\mathrm{CMB} E_e^2}{3m_e^2c^4} \simeq 3.24\left(\frac{E_e}{1\;\mathrm{TeV}}\right)^2 \mathrm{GeV}
\end{equation}
where $\epsilon_\mathrm{CMB} = 6.34\times 10^{-4}\;\mathrm{eV}$ is the mean energy of a CMB photon, $c$ is the speed of light in vacuum, $m_e$ is the mass of an electron. Leptons of $10\;\mathrm{TeV}$ give secondary photons in the energy range of the CTAO, at $300\;\mathrm{GeV}$.

In the Thomson regime ($E\lesssim 30\;\mathrm{TeV}$), the energy lost by a lepton at each interaction is small and the energy decrease is continuous:
\begin{equation}
\label{loss}
    E_e(x) = \frac{E_e^0}{1+x/D_\mathrm{ic}^0}
\end{equation}
where $x$ is the travelled distance from the pair creation site, $E^0_e$ is the initial lepton energy, $D_\mathrm{ic}^0=(3/4)(m_e c^2)^2(\sigma_T \rho_\mathrm{CMB}E_e^0)^{-1}$ is the initial cooling distance due to Compton scattering, $\sigma_T$ is the Thomson scattering cross section and $\rho_\mathrm{CMB}=0.26\;\mathrm{eV}\;\mathrm{cm^{-3}}$ is the energy density of the CMB (see \citealp{PhysRevD}). A lepton with $E_e^0=10\;\mathrm{TeV}$ loses merely $10\%$ of its energy after travelling  $x\sim 1\;\mathrm{kpc}$. 

The two leptons are deflected by the IGMF as illustrated in Fig. \ref{Triangle}. This deflection angle $\delta$ is characterised by the Larmor radius:
\begin{equation}
    R_L = \frac{E_e}{qB}
\end{equation}
where $q$ is the lepton charge. For large correlation lengths, $\lambda_B\gg D^0_{\mathrm{ic}}$, the lepton up-scatters most of its target photons in a region where the IGMF can be considered as uniform. After travelling a distance $x$, the total deflection is simply the sum of infinitesimal deflections:
\begin{equation}
    \delta(x) = \int^x_0 \frac{c\mathrm{d}t}{R_L(t)}\, .
\end{equation}
It follows that the magnetic deflection takes the form:
\begin{equation}
\label{deflection_Mpc}
    \delta=\frac{D^0_{ic}}{2R^0_L}\left[ \left(\frac{E^0_e}{E_e}\right)^2-1\right] \approx \frac{D_{ic}}{2R_L}\, . 
\end{equation}
In this formula, $R^0_L$ is the initial Larmor radius and $D_\mathrm{ic}(x)$ is the cooling distance of the lepton after several Compton scatterings. 

Most GRBs are observed at distances $D\sim 1 \;\mathrm{Gpc}$ (redshift exceeding $0.1$), which are much larger than the original gamma-ray absorption length $\lambda_{\gamma\gamma}$. Since $\sin\theta = (\lambda_{\gamma\gamma}/D)\sin\delta$ and $\theta \ll 1$, the time delay $\Delta t$ simplifies to:
\begin{equation}
\label{triangle_formula}
    c\Delta t \approx \frac{\lambda_{\gamma\gamma}}{2}\delta^2\, .
\end{equation}
From combining equations \ref{compton} to \ref{triangle_formula}, the average energy of a secondary photon is therefore:
\begin{equation}
\label{energy}
    \langle E_\gamma \rangle \approx 1\left(\frac{B}{10^{-17}\;\mathrm{G}}\right)\left(\frac{\lambda_{\gamma\gamma}}{100\;\mathrm{Mpc}}\right)^{1/2}\left(\frac{\Delta t}{2\;\mathrm{days}}\right)^{-1/2}\mathrm{TeV}
\end{equation}
Eq. \ref{energy} implies that, in a observation time window corresponding to a given time delay $\Delta t$, the average energy at which the cascade emission peaks depends sensitively on the field strength: for weak fields, the deflection of pairs is minimal, causing the cascade to peak at lower energies. In contrast, stronger fields lead to larger deflections, which shifts the cascade peak to higher energies, sometimes into the $\mathrm{TeV}$ range. At these energies, partial reabsorption due to interactions with the EBL can occur, making the spectral behaviour more complex.

From Eq. \ref{energy} we can estimate the constraints accessible to an instrument sensitive in a given energy range. Assuming observations of several weeks, Fermi-LAT energies (about $10\;\mathrm{GeV}$) allow covering strengths up to several $10^{-19}\;\mathrm{G}$ as derived by various authors (\citealp{Vovk_2023a}, \citealp{Huang_2023}, \citealp{Dzhatdoev_2023}). In contrast, IACTs have access to larger energies, from a few tens of $\mathrm{GeV}$ to a few tens of $\mathrm{TeV}$, and they can probe magnetic field strengths up to $ 10^{-17}\;\mathrm{G}$ provided they observe GRBs for long enough (several nights). 

The estimations derived so far have assumed a large correlation length. When the correlation length is smaller than the lepton mean free path, $\lambda_B \lesssim 10\;\mathrm{kpc}$,  the particle travels through regions of different magnetic properties before it cools down significantly. In this diffusive regime, the magnetic deflection can be approximated by a random walk with a magnetic deflection reduced by a factor $\sqrt{x/\lambda_B}$. To the first order, for the same time delay and magnetic field, the cascade peaks at a lower energy with a higher flux. Consequently, stronger magnetic fields are detectable at smaller correlation lengths, as we will see later.

\subsection{Numerical simulations}

The results presented in this work are based on a recent version of a Monte Carlo simulation code, \verb|CascadEl|\footnote{\url{https://gitlab.com/rbelmont/cascadel2}}, first presented in \cite{Fitoussi_2017_bis}. It is a full 3D Monte Carlo code tracking photons and leptons during their travel to Earth. 

It includes interactions with background photons through photon-photon pair production and Compton scattering. The background photon field is composed of the CMB and EBL. The results presented here assume the model by \cite{dominguez2011}. The interactions are computed with any photon from this total background field, whether it comes from the EBL or the CMB. All interactions are computed according to the exact first-order cross sections in QED. In particular, the scattering of high-energy leptons is described by the Klein-Nishina cross section. At each interaction, the energy and direction of the background photon are drawn randomly according to probability functions corresponding to the cross sections. The interaction outcome is also computed according to similar probabilities. This includes not only the energy of the secondary particles but also their direction. In the physical process, the secondary particles are produced in directions that are slightly different from the direction of the parent high-energy particle. This induces a weak intrinsic spreading of the cascade, which can dominate the magnetic field spreading for weak magnetic field strengths. 

Simulations are performed in a full cosmological frame, including effects of expansion on length and time scales, on magnetic strength, and on photon and lepton energies. They also include the cosmological evolution of the background photon field, assuming a $\Lambda\mathrm{CDM}$ cosmological model with $H_0=67.4 \;\mathrm{km}\cdot \mathrm{s}^{-1}\;\mathrm{Mpc}^{-1}$, $\Omega_m=0.3$ and $\Omega_\lambda=0.7$ (\citealp{PhysRevD}). 

The photons propagate in straight lines and the leptons are deflected by the IGMF, modelled as a set of cubic cells of size $\lambda_B$ in which the field direction is assumed constant. The field amplitude is fixed, but its orientation varies randomly from one cell to another. In each cell, the lepton trajectory is helical\footnote{with a suitable choice of cosmological coordinates}, which allows for an exact and efficient computation without the need for a numerical integrator. For large correlation lengths,  leptons can visit only one cell before they cool down. In that case, the result depends on the specific orientation in that unique cell, and therefore on the specific realisation of the magnetic structure. To obtain mean properties not attached to a specific direction, all results presented here have been averaged over $100$ realisations of the magnetic structure. 

In electromagnetic cascades, the number of particles decreases rapidly with increasing energy, but the high-energy component is the most relevant for our study. As a result, the Monte Carlo statistics is highly unbalanced, with most of the computational time spent tracking low-energy particles. To optimise computational time, several measures are taken. First, energy thresholds are used for leptons and photons to prevent tracking particles below any relevant energy range. Second, low-energy leptons are allowed to travel several Compton distances before proceeding to a numerical scattering. Indeed, Compton cooling becomes less efficient at low energy, meaning that neighbour secondary photons have similar properties. It is therefore more efficient to compute only one interaction every several interaction lengths and to give the produced photons a larger weight. Third, the code discards a fraction of the produced particles at each interaction, while the weight of the remaining particles is increased accordingly. \\

A simulation is run for given magnetic field strength $B$,  correlation length $\lambda_B$ and source redshift $z$. The primary photons are emitted with random energies following an ad hoc power-law spectrum, at a given time and in a unique direction. Primary and secondary particles are tracked until they cross the \textit{observer sphere}, i.e. a sphere centred on the source and with comoving radius the comoving distance to the source at redshift $z$. At the observer sphere, the following properties are recorded:
\begin{itemize}
\item The energy $E_i$ of the recorded particle at the observer sphere.
\item The time delay $\Delta t_i$ compared to a primary photon reaching the observer sphere in a straight line.
\item The spherical coordinates on the observer sphere relative to the emission direction of the primary photons.
\item The two angles describing the observation direction, i.e. the propagation direction relative to the source-detection axis. 
\item The photon generation (starting with $0$ for the primary photons).
\item The energy $E^0_i$ of the primary photon.
\item The particle weight $w_i$ due to the particle and Compton samplings described above. 
\end{itemize}

The code has been checked against \verb|CRPropa|\footnote{\url{https://github.com/CRPropa}}, a numerical framework originally developed for modelling the propagation of cosmic rays (\citealp{Alves_Batista_2022}). Both codes give similar results in several situations relevant to the present work. The slight deviations are due to the approximations on the cosmic evolution of the EBL in \verb|CRPropa| (see discussion in \citealp{Kalashev_2023}). In comparison, \verb|CascadEl| incorporates the predicted photon density of a given EBL model, and generates interactions with Monte Carlo simulations.

\subsection{Post processing}
\label{processing}

From the events produced by one such simulation at a given ($z,B,\lambda_B$), it is possible to derive, as a second step, the secondary emission produced by sources with any spectral, time and angular properties. The code used to analyse the results of the Monte Carlo simulation has been gathered in a Python package, \verb|Cascapy|\footnote{\url{https://gitlab.com/rbelmont/cascapy}}.

At start, the events are reweighed according to the energy $E^0_i$ of their ancestor, the primary photon, to match the desired source spectrum.  The events are also reweighed by the recorded angles to reproduce the cascade induced by a source of any angular distribution (e.g. a jet of given opening angle), and to model the PSF of any instrument. On the short time scales of GRB observations, all secondary gamma rays are observed well inside the PSF of any instrument, and were produced by primaries emitted only very close to the line of sight, meaning that the exact angular distribution is irrelevant. For these reasons, the results presented here are integrated over the entire sky  assuming that the primary source emits isotropically. In principle, this angular re-weighting can only be applied by assuming the universe around the source is isotropic, meaning that primaries emitted in any direction around the source produce similar cascades. As one single realisation of the IGMF structure is not isotropic at the smallest scales, this assumption can be debated for large correlation lengths (larger than $10-100\;\mathrm{Mpc}$) when producing images of individual sources. However, we are interested only in spatially integrated fluxes. Moreover, this work assumes  correlation lengths of $1\;\mathrm{Mpc}$ at most, and the results are averaged over many magnetic field realisations. In such cases, the reconstruction of any property of the intrinsic primary emission is accurate. 

Finally, the events can be reweighed to account for any temporal evolution of the primary emission. Assuming a primary flux $F_0(t_0)$ as it would be observed at time $t_0$ if there were no interaction with the EBL, the observed flux averaged between times $t_1$ and $t_2$ is derived from the following correlation:
\begin{equation}
F(t_1,t_2)  = \frac{1}{t_2-t_1} \int_{t_1}^{t_2} \mathrm{d}t \int_{-\infty}^{t} dt_0 F_0(t_0) K(t-t_0) \\
\end{equation}
where the cascade kernel $K(t-t_0)$ is a function of the time delay $\Delta t = t-t_0$. This correlation can also be written: 
\begin{equation}
F(t_1,t_2) =   \int_{0}^{\infty} a(t_1,t_2,\Delta t) K(\Delta t)  \mathrm{d}(\Delta t) 
\end{equation}
with
\begin{equation}
 a(t_1,t_2,\Delta t) = \frac{1}{t_2-t_1} \int_{t_1-\Delta t}^{t_2-\Delta t} \mathrm{d}t_0 F_0(t_0)  
\end{equation}
When applied to results of Monte Carlo simulations, a given flux of primary emission is hence produced by re-weighting each event $i$ with the factor $a(t_1,t_2, \Delta t_i)$. Complex models of primary GRB evolution require numerical integration of this factor and can be quite time demanding when dealing with millions of events. However, simple models (e.g. power laws) allow for fast, analytical integration.

Previous works have mostly modelled the GRB emission as a single pulse at the explosion time (\citealp{Dzhatdoev_2020}, \citealp{Huang_2023}, \citealp{Xia_2024}). This pulse approximation is acceptable for sources observed a long time after the primary flux has faded away. However IACT observations are usually limited to a few nights as the VHE signal rapidly fades away. On these time scales the primary emission still contaminates the secondary emission. Moreover the primary emission keeps producing new secondary photons even at late time. Such late contributions produced short before the observation time correspond to shorter time delays and are expected at higher energies. Therefore, using an explicit model for the time evolution of the primary emission can be crucial for transient sources such as GRBs when observed with IACTs.

\begin{figure}[h!]
    \centering
    \includegraphics[width=1\linewidth]{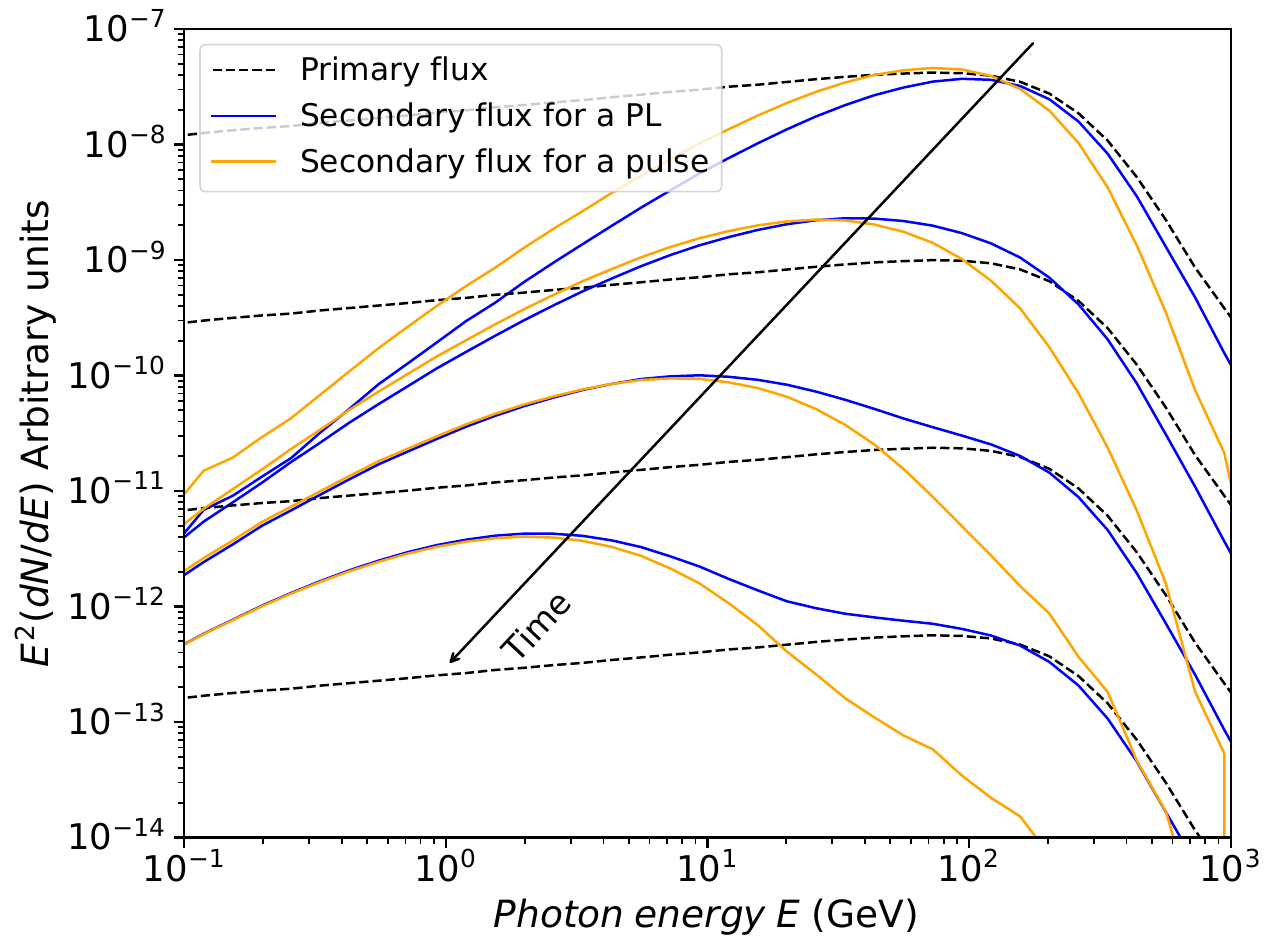}
    \caption{Spectra of a fiducial GRB following a power law in energy with spectral index $\gamma=1.8$ and with $(B,\lambda_B)=(10^{-19}\;\mathrm{G},1\;\mathrm{Mpc})$. The orange and blue lines show the secondary emissions assuming respectively a pulse and a power-law decay ($\alpha=1.2$) of primary emission. The primary emission in the latter case is shown as dashed, black lines. The four sets of curves correspond to the flux integrated on the following time periods from top to bottom: $[10\mathrm{s}-4\;\mathrm{min}]$, $[4\;\mathrm{min}-1.5\;\mathrm{h}]$, $[1.5\;\mathrm{h}-30\;\mathrm{h}]$ and $[30\;\mathrm{h}-1\;\mathrm{month}]$}
    \label{PulsePL}
\end{figure}

Fig. \ref{PulsePL} illustrates the role of the late afterglow in the emission from a generic GRB at $z=0.4$. It shows the total spectra in $4$ logarithmic time windows spanning the first month of observation ($[10\mathrm{s}-4\;\mathrm{min}]$, $[4\;\mathrm{min}-1.5\;\mathrm{h}]$, $[1.5\;\mathrm{h}-30\;\mathrm{h}]$ and $[30\;\mathrm{h}-1\;\mathrm{month}]$). The primary spectrum is assumed to be a power law with index $\gamma= 1.8$ and maximal energy $E_\mathrm{max} = 30\;\mathrm{TeV}$, and the IGMF has $B=10^{-19}\;\mathrm{G}$ and $\lambda_B=1\;\mathrm{Mpc}$. The orange line shows the secondary emission computed assuming that the total fluence is emitted as a single pulse in time. This secondary emission follows the expected evolution, namely a decreasing flux moving to lower energy (Eq. \ref{energy}). In comparison, the blue and black lines show the secondary and primary emissions assuming the photon flux decays following a power-law (with index $\alpha = 1.2$, initial time $t = 6\mathrm{s}$ and total fluence over $1$ month identical to the pulse fluence). Because the temporal decay is rapid, most of the GRB fluence is emitted during the first minutes of the burst, producing a secondary  very similar in shape and amplitude to the pulse case, which quickly falls to the Fermi energy band. However, the primary photons emitted at the late times of the afterglow produce an excess hard tail of secondary emission, mostly visible in the VHE band, a feature particularly relevant to IACT observations. Its spectral slope and amplitude can vary significantly depending on the time evolution and intrinsic spectral shape of the source. For a power-law decay, this excess secondary emission also decreases as a power law in time at the latest times, with the same index (i.e. with a constant amplitude with respect to the primary flux).

\subsection{Synthetic data}

To bridge theoretical models with expected observations, this section details the generation of realistic, synthetic CTAO data sets for four historical VHE GRBs: GRB 180720B (\citealp{Abdalla_2019}), GRB 190829A (\citealp{Hess_2021}), GRB 190114C (\citealp{2019}), and GRB 221009A (\citealp{Lhaaso_2023}). Using the previously described procedure, we construct these data sets by extracting their published spectral and light curve properties, combining them with IGMF cascade simulations, and subsequently folding the resulting fluxes through a realistic CTAO response.

From the respective publications, we extract the parameters to model the primary emission as the product of two power laws\footnote{As en exception, the intrinsic lightcurve of GRB 221009A is modelled with a piecewise power law constituted of 5 intervals.} characterised by a cut-off energy $E_\mathrm{cut}$, an energy index  $\gamma$ and a time index $\alpha$, from time $t_{min}$, accounting for the delay required for the afterglow to develop (see Table \ref{anymode}). The differential photon flux of primary emission is: 
\begin{equation}
\label{model}
    \Phi(E,t>t_\mathrm{min}) = \Phi_0  E^{-\gamma}t^{-\alpha}\exp \;(-E/E_\mathrm{cut}) \quad \mathrm{(in\; ph/erg/s)}.
\end{equation}
Due to the EBL absorption, the intrinsic cut-off energy is poorly constrained in GRB observations. Regarding our synthetic data, we model every GRB afterglow emission with $E_\mathrm{cut} = 10\;\mathrm{TeV}$, unless otherwise specified. This value is consistent with the highest energy photons detected by LHAASO from GRB 221009A ($13 \; \mathrm{TeV}$, \citealp{cao_2023}). As we will discuss later, this choice of $E_\mathrm{cut}$ has a major impact on the results, as the secondary photons originate from the most energetic part of the spectrum. 

\begin{table*}
\centering
\caption{Main parameters of the intrinsic unabsorbed models, used to generate the synthetic data. For GRB 221009A (\citealp{Lhaaso_2023}), LHAASO identified five temporal indices corresponding to time periods defined by $t_i = [0,\;4.85,\;15.37,\;22,\;670,+\infty]\;\mathrm{s}$. Only the index of the last interval is shown in the table. $E_{EBL}$ indicates the energy at which half of the primary flux is absorbed through interactions on the EBL.}
\label{anymode}
\begin{tabular}{c c c c c}
\hline\hline 
$\mathrm{GRB}$                        & \thead{180720B}& \thead{190114C}& \thead{190829A}& \thead{221009A}\\    
\hline
$E_\mathrm{iso}\;[\mathrm{erg}]$&  $6\times 10^{53}$             & $2.5\times 10^{53}$             & $2\times10^{50}$              &  $1\times 10^{55}$   \\[2pt]
Redshift $z$                         &  0.653         &  0.425        &   0.08         &  0.151\\[2pt]
$E_\mathrm{EBL}\;[\mathrm{GeV}]$      &  142           &  201          &  811           &  433\\[2pt] 
$\mathrm{Spectral \; index}\; \gamma$ & 1.60           &  2.22         &  2.06          &  2.32\\[2pt] 
$\mathrm{Temporal \; index}\; \alpha$ &  1.00          &  1.60         &  1.09          &  2.21\\[2pt] 
$\mathrm{Flux}\; \Phi\;(1 \; \mathrm{TeV}, 1\; \mathrm{h}) \; \left[\mathrm{TeV}^{-1}\;\mathrm{s}^{-1}\;\mathrm{cm}^{-2}\right]$
                                      &  $3.79\times 10^{-10}$ &  $6.10\times 10^{-11}$ &    $4.65\times 10^{-11}$ &  $1.94\times 10^{-9}$\\[2pt] 
$\mathrm{VHE\;start}\; t_\mathrm{min} \; [\mathrm{s}] $ , this work
                                      &  $100$         &  $6$          &  $1600$        &  $15.37$\\[2pt]
VHE Observatory                       & H.E.S.S & MAGIC & H.E.S.S & LHAASO / CTAO \\[2pt]
Delay before detection                & $10\mathrm{h}$     & $62\mathrm{s}$    & $4\mathrm{h}20$    & $225\mathrm{s}$ \\ 
  \hline
\end{tabular}
\end{table*}
These models of time and spectral evolution are applied to the results of Monte Carlo simulations at the GRB redshift to produce the total emission (including both the primary and secondary emissions) arriving on Earth at any observation time, as explained in Sec. \ref{processing}. Synthetic data sets are then created from the CTAO response simulated using the \verb|Gammapy| software (\citealp{Donath_2023}, \citealp{gammapy:zenodo-1.2}). The released CTAO prod5 Instrument Response Functions (IRFs) (\citealp{cherenkov_telescope_array_observatory_2021_5499840}) for the Alpha configuration (see Sec.\ref{Intro}) are used. The IRFs comprise the response of the telescope array and are computed at fixed zenith angles ($20^\circ$, $40^\circ$, $60^\circ$) assuming optimal and stable observing conditions (nights without Moon, no clouds). An internal study reveals an energy resolution of $25\%$ at low energies, allowing the derivation of the following reconstructed energy thresholds : $30\;\mathrm{GeV}$, $40\;\mathrm{GeV}$, and $110\;\mathrm{GeV}$ in the North, and $60\;\mathrm{GeV}$, $110\;\mathrm{GeV}$, and $350\;\mathrm{GeV}$ in the South, respectively for each zenith angle. The IRFs are made available for durations of $30'$, $5h$ and $50h$. 

For each GRB, the altitude evolution and the nighttime periods are computed from its galactic coordinates and the date of the external alert trigger, using the \verb|Astroplan| package (\citealp{Morris_2018}). For the computation of the visibility windows, we choose to ignore the presence of the Moon, although it affected some of the published observations. As an approximation, we still check the robustness of our results under partial Moonlight.  The GRB is assumed undetectable below an altitude $24^\circ$, where the IRFs become unreliable, although some published observations were made below that angle.
The flux in a night period is sampled at logarithmically spaced times, creating intervals from about $10 \mathrm{s}$ up to a full night in duration, to have sufficient count statistics at any time. The relevant IRF is attributed to each valid observation time interval based on the altitude\footnote{We checked that defining the altitude at the start, the end or the center of the time bin yields similar results.} and duration of the interval, thus allowing the computation of the corresponding signal and background in the field-of-view using a reflected region technique (\citealp{Berge_2007}) . 

\begin{figure}
    \centering
    \includegraphics[width=1\linewidth]{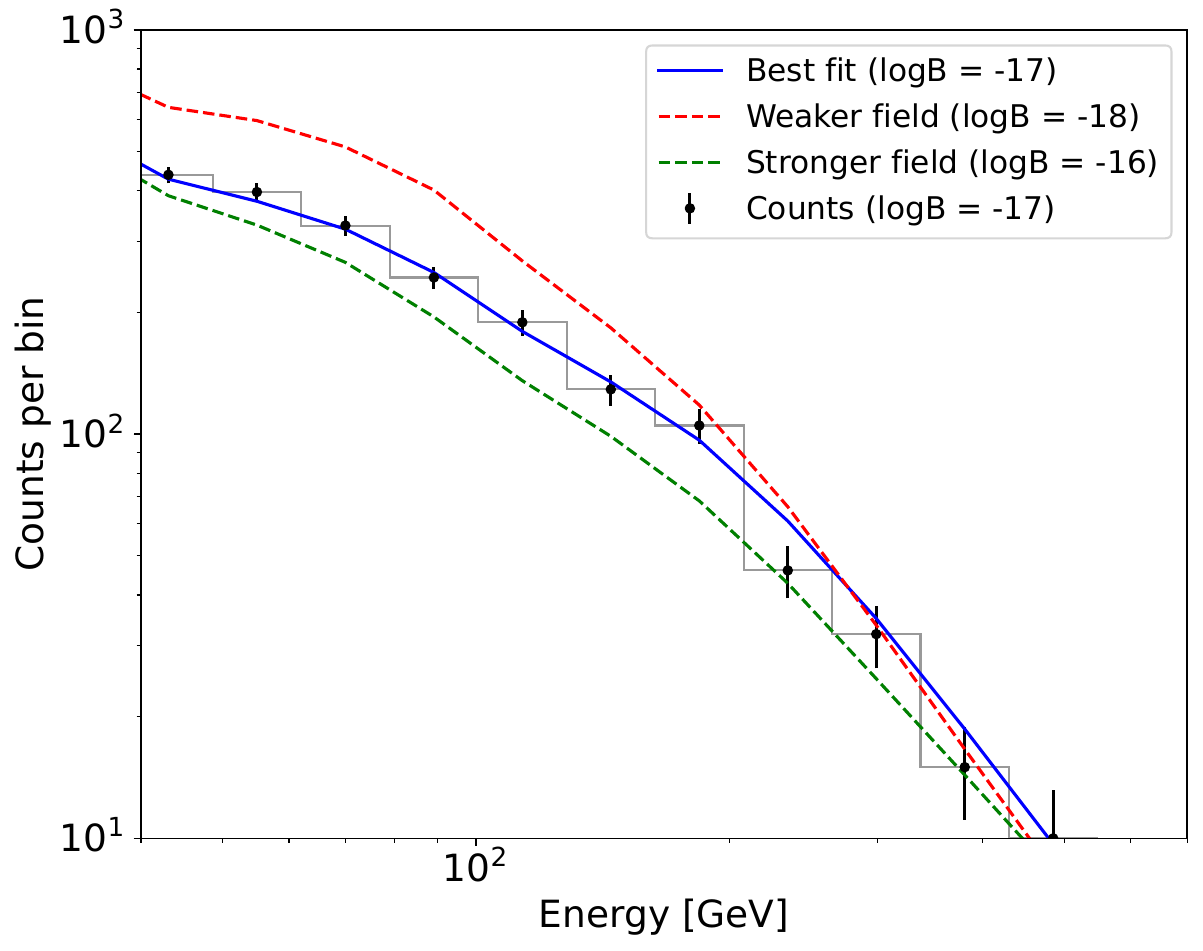}
    \caption{Example of signal counts in the North site (LSTs plus MSTs) at the end of the first night of GRB 190114C (2.29h - 2.61h for data generated with $B=10^{-17}\;\mathrm{G}$. The best fit (blue) at $10^{-17}\;\mathrm{G}$ is compared with two model fits at $10^{-18}$ (red) and $10^{-16}\;\mathrm{G}$ (green).}
    \label{on-counts}
\end{figure}

\subsection{Fit method and analysis}

Previous studies analysing the imprints of the IGMF generally follow a two-step approach. First, the intrinsic GRB properties are inferred by fitting the very early afterglow, assuming that it is not contaminated by secondary emission. The cascade emission is then derived from these intrinsic properties. This approach, however, fails to consistently propagate uncertainties from the primary flux to the characteristics of the secondary emission and to address the possible degeneracy between primary and secondary emissions. 

In contrast, the present analysis jointly accounts for the spectral and temporal properties of the cascade, allowing the intrinsic properties of the primary emission and the IGMF amplitude to be inferred within a single framework. Namely, data are fitted with a table model for the total emission computed assuming an intrinsic emission described by Eq. \ref{model}. This model has five free parameters: the intrinsic time index $\alpha$, spectral index $\gamma$, cut-off energy $E_{\mathrm cut}$ , normalisation $\eta$, and the logarithm of the IGMF strength $log(B)$. The correlation length is fixed at $1 \; \mathrm{Mpc}$ for all GRBs\footnote{plus $1\;\mathrm{kpc}$ and $1\;\mathrm{pc}$ for GRB 190114C, see Sec. \ref{GRB190114C_lambda_B}}. Allowing $\lambda_B$ to be free makes the fit difficult to converge. In particular, the effect of $\lambda_B$ in the random walk regime ($\lambda_B \lesssim 10\;\mathrm{kpc}$) is partly degenerated with $B$. The model interpolates in a table built from results of Monte Carlo simulations for parameter values in grids defined as in Table \ref{dimensions}. In this study, a different table is used for each GRB, incorporating the appropriate redshift and observation windows. The latter depend on the source altitude and the nights duration. For each parameter set, the table contains fluxes for $100$ logarithmically-spaced bins in energy, from $1 \;\mathrm{GeV}$ to $20 \;\mathrm{TeV}$. 
\begin{table}
\caption{Limits and number of values for the table model used to fit the data. The grids are linear for the two indices, and logarithmic for the cut-off energy and magnetic strength.} 
\label{dimensions}      
\centering                                      
\begin{tabular}{c c c c}          
\hline\hline                        
$\mathrm{GRB}$& \thead{Min}& \thead{Max}& \thead{Number of values}\\    
\hline
$B\; (\mathrm{Magnetic \; strength})$& \thead{$10^{-21}\;\mathrm{G}$}& \thead{$10^{-12}\;\mathrm{G}$}&
  \thead{23}\\[2pt]
$\gamma\; (\mathrm{Spectral \; index})$&\thead{1}& \thead{3}&
  \thead{8}\\[2pt]
$\alpha \; (\mathrm{Temporal \;index})$& \thead{0.8}& \thead{2.8}&
  \thead{8}\\[2pt]
$E_\mathrm{cut} \; (\mathrm{cut-off \; energy}) $& \thead{$1\;\mathrm{TeV}$}& \thead{$70\;\mathrm{TeV}$}&
  \thead{20}\\[2pt]
\hline                                             
\end{tabular}
\end{table}
These count numbers are randomised once to mimic real observation datasets. The fit uses the \verb|Gammapy| library (\citealp{gammapy:zenodo-1.2}) to disentangle the gamma-ray contribution from the residual cosmic-ray background on the data of the two sites, separately and/or jointly. 

As discussed later, the maximal energy of the intrinsic spectrum $E_{\rm cut}$ and the IGMF strength $B$ are partly degenerate, which sometimes makes convergence of the fit difficult, with many local minima. To ensure robust and reliable convergence across all analysed cases, we apply the following fitting procedure:
\begin{enumerate}
    \item For a grid of fixed $E_{\rm cut}$ and $B$ values, we first fit the $3$ other parameters, ($\alpha, \gamma,  \eta$), and get their values at the best fit, ($\alpha_1, \gamma_1,  \eta_1$).
    \item Then, for each $B$ value, we find the value $E_{\rm cut, 1}$ of the best fit among all tested cut-off energies, using ($\alpha_1, \gamma_1,  \eta_1$, $E_{\rm cut, 1}$) as the starting values for new fits, letting the cut-off energy free. The best fit returns ($\alpha_2, \gamma_2,  \eta_2$, $E_{cut,2}$) for each $B$ value. This step is also used to define the confidence levels for the magnetic field.
    \item Last, we find the best magnetic field $B_1$ among all tested values, and we perform a global fit starting from ($\alpha_2, \gamma_2,  \eta_2$, $E_{cut,2}$ , $B_1$) with $B$ as a free parameter, as well as all the other parameters.
\end{enumerate}

When the count numbers are sufficiently high, the fluctuations are small, and unsurprisingly, the fit converges to the B value of the generated data. As an illustration, Figure \ref{on-counts} shows the binned dataset counts at the end of the first night for GRB 190114C, assuming $B=10^{-17}\;\mathrm{G}$, together with the fitted models for $B=10^{-18}\;\mathrm{G}$ and $B=10^{-19}\;\mathrm{G}$.

For an assumed value of the magnetic field, the fits are performed on a single dataset corresponding to a single randomisation of the underlying model. However, the time intervals are long enough to have more than $10$ counts in each energy bin and the signal spreads over many bins, so that generating more than one trial does not significantly change the results.

\begin{figure}
    \centering
    \includegraphics[width=1\linewidth]{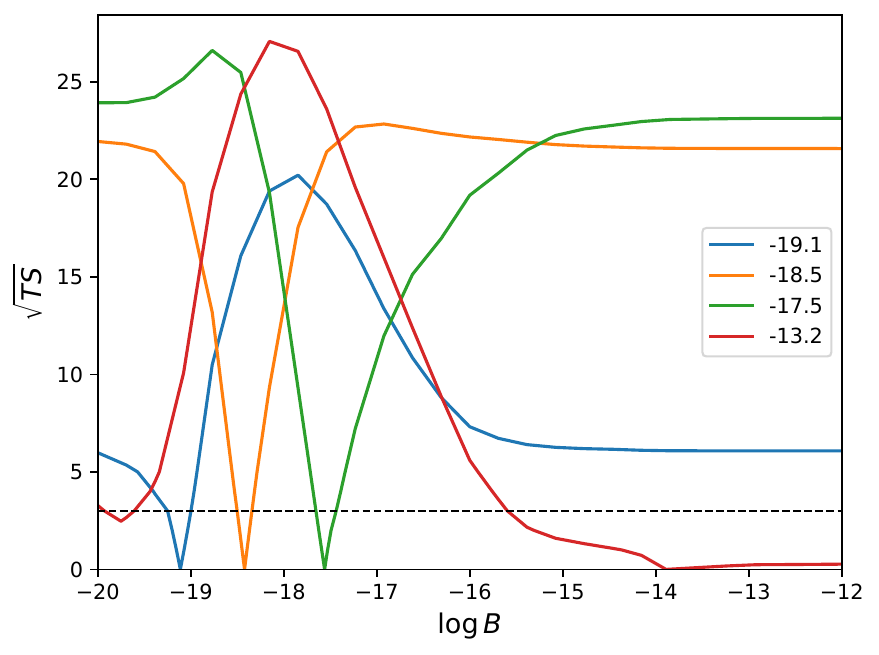}
    \caption{Likelihood profiles for four values of $\log B$. The dashed line indicates the $3\sigma$ confidence level for GRB 190114C.}
    \label{profiles}
\end{figure}

To derive these parameters, the fit relies on the \texttt{Minuit} algorithm (\citealp{James_1975}) to maximise the log-likelihood estimator. Specifically, we use the \texttt{WStat} statistic implemented in \texttt{Gammapy}\footnote{\url{https://docs.gammapy.org/1.2/user-guide/stats/fit_statistics.html}}, which computes the likelihood $\mathcal{L}$ based on the expected and measured counts in the source (ON) and background (OFF) regions (see \citealp{lima_1983}). The fit result consists of the best set of parameters, in particular, the magnetic field strength $B_{best}$. The statistical uncertainties for this parameter are derived from the likelihood profile that compares the best likelihood at a given $B$ (with all other parameters free) to the likelihood at $B_{best}$, namely the quantity $\sqrt{\Delta TS(B_{best}, B)}$. The test statistics difference $\Delta \mathrm{TS}$ is written as: 
\begin{align}
\label{variance}
  \Delta TS(B_{best}, B)= - 2 \log \frac{\mathcal{L}(B_{best})}{\mathcal{L}(B)}
\end{align}
Examples of such likelihood profiles are shown in Fig. \ref{profiles}. For each GRB analysed, synthetic datasets are generated over a range of IGMF strengths are subsequently analysed using this fit procedure to assess the CTAO ability to constrain the IGMF. As we will see in Sec. \ref{spectral-temporal}, the result consists of likelihood maps where the likelihood has been scanned over a range of IGMF strengths (e.g. Fig. \ref{FitMap_GRB19}). Colours indicate the deviation from the best fit. Although plotted in a single map, each vertical profile is independent of the others since it corresponds to a different dataset and a different reference likelihood ${\mathcal{L}(B_{best})}$. 

When the count number is low and the B-induced distortions are small the fit does not always converge and the magnetic strength gets blocked at the boundary $B=10^{-12}\;\mathrm{G}$ (see Fig. \ref{FitMap_GRB19}). However, in such cases, the likelihood profile is essentially flat at the boundary and extending the table would not change the lower limit on the IGMF.

\section{Detected very high-energy GRBs in CTAO}

In this section, we apply the previously described methodology to four gamma-ray bursts detected in the $\mathrm{TeV}$ range. We begin with a detailed analysis of GRB 190114C, which occurred at night, above the horizon, i.e. without delay in detection, and was among the brightest GRBs. To explore the influence of spectral and temporal characteristics, as well as observation conditions, we repeat this analysis for three additional sources: GRB 221009A, GRB 180720B and GRB 190829A. For GRB 221009A, we also examine existing data from the CTAO LST-1 to compare current observational capabilities with the expected performance of the full array.

\subsection{GRB 190114C}
\label{GRB190114C}
On January $14^\mathrm{th}$, 2019, 20:57 (universal time), denoted $T_0$ in what follows, GRB 190114C was simultaneously detected by the Gamma-ray Burst Monitor (GBM) onboard the Fermi satellite and the Burst Alert Telescope (BAT) on the Swift satellite at $\mathrm{RA}=54.51^\circ$, $\mathrm{DEC}=-26.94^\circ$ (\citealp{Ajello_2020}). The Nordic Optical Telescope measured a redshift of $z=0.425$ (\citealp{Selsing_2019a}). The isotropic-equivalent energy is estimated to be $2.5\times 10^{53}\;\mathrm{erg}$ in the Fermi energy band (\citealp{2019_bis}). MAGIC observed the source between $100\;\mathrm{GeV}$ and $10\;\mathrm{TeV}$, starting at $T_0+62\mathrm{s}$ and reaching a maximal significance of $51\sigma$ at $T_0+19\;\mathrm{min}$, despite degraded observation conditions due to the presence of Moonlight and low altitudes (\citealp{2019}). The MAGIC data are well fitted in energy and time as independent power laws with respective indices $\gamma=2.22 \ ^{+0.23} _{-0.22}$ (from $T_0+62\mathrm{s}$ to $\sim T_0 + 2400\mathrm{s}$, ) and $\alpha=1.60 \ \pm \ 0.07$ (from $0.3$ to $1 \;\mathrm{TeV}$). The simulated data are reweighed according to the model of Eq. \ref{model} with the central measured values. We follow the assumption of \cite{ravasio_2019} that the afterglow started at $t_\mathrm{min}=T_0+6\mathrm{s}$, as detected by \textit{Fermi}-LAT (\citealp{Ajello_2020}). The flux normalisation, $\Phi_0$, is obtained from  $ \frac{1}{2454-62}\int_{62}^{2454}\Phi(E,t)\;\mathrm{d}t = 8.45\times 10^{-9}\;\mathrm{TeV}^{-1}\;\mathrm{s}^{-1}\;\mathrm{cm}^{-2}$ at $E = 460\;\mathrm{GeV}$, as in MAGIC (\citealp{2019}). 

The burst occurred during the night, and, as MAGIC did, we assume a $60\mathrm{s}$ delay before the observation starts, comprising a Swift alert delay ($22\;\mathrm{s}$) and the LST slewing time ($30\mathrm{s}$ at most). We use the full Array IRFs from the start, neglecting the longer slewing time of MSTs. During all the observation periods, lasting $3$ and $4$ hours for the first and subsequent nights respectively, the source altitude never exceeded $30^\circ$, corresponding to an effective $110\;\mathrm{GeV}$ detection threshold. In the South, the first night started $2$ hours after $T_0$. During this five-hour observation the source was at higher altitudes, allowing a lower $60\;\mathrm{GeV}$ energy threshold despite the absence of LSTs. Under these assumptions, accumulating the signal and background counts over the time slices from the accurate IRFs in each slice, the GRB is detected with a cumulated significance of $1388\sigma$ in the North after $30$ minutes of observation, and $67\sigma$ in the South at $T_0+9$ hours.
 
\subsubsection{Semi-quantitative approach}

In the following section, we evaluate the detectability of IGMF signatures by modelling the primary and secondary emissions of GRB 190114C. By comparing these models to the anticipated sensitivity of CTAO, we obtain first indications for the specific time windows, energy ranges and magnetic field strengths at which the delayed cascade flux becomes observable.
\begin{figure}
    \centering
    \includegraphics[width=1\linewidth]{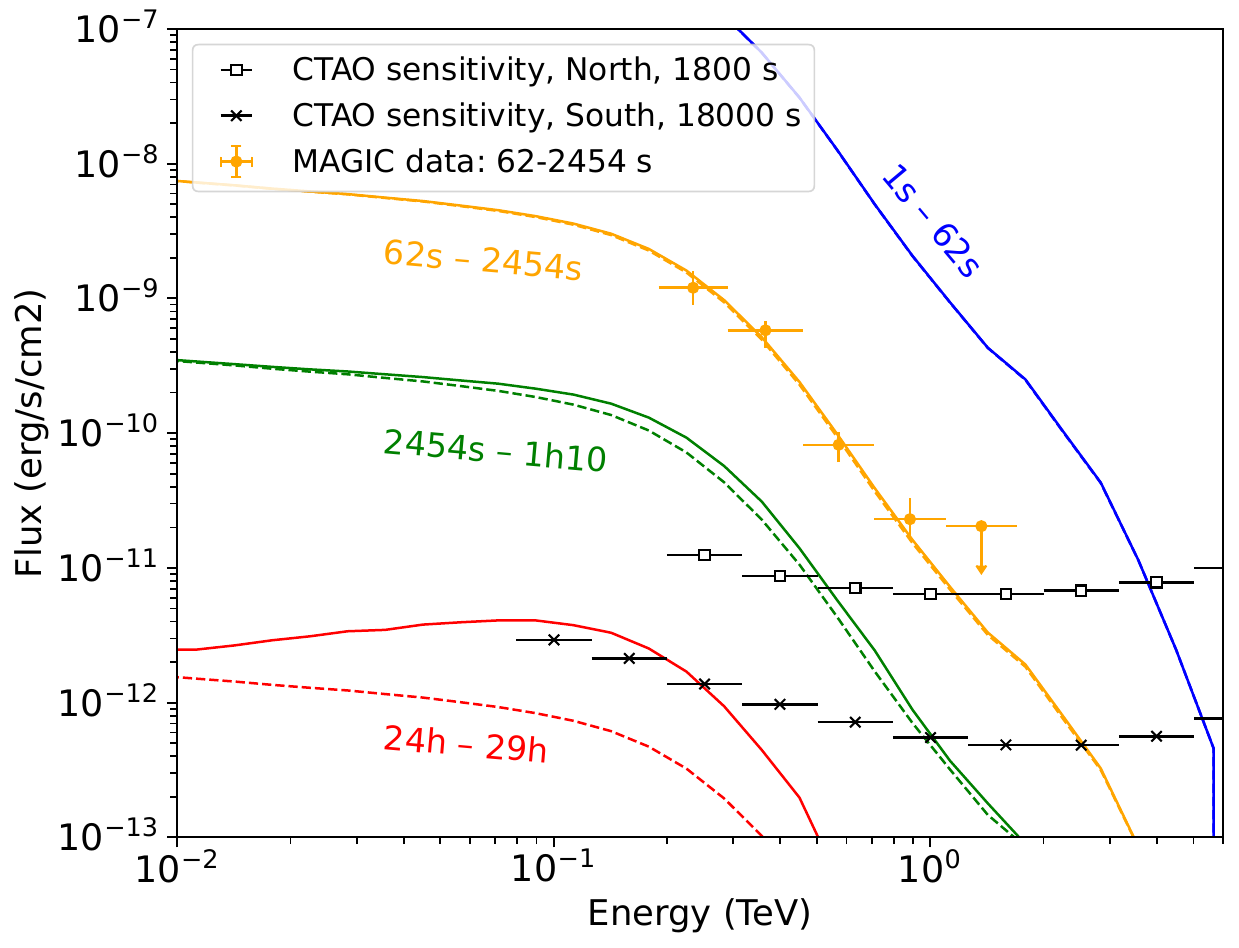}
    \caption{Averaged spectra of GRB 190114C for several time ranges with a magnetic strength of $10^{-17}\;\mathrm{G}$. The total and primary spectra are shown with dashed and solid lines, respectively. The MAGIC data are shown for the second period. The red lines are for the second night. The black points show the CTAO sensitivity curve, for North (black squares, 30-minute observation, $60^\circ$ of zenith angle) and South (black crosses, five-hour observation and a $20^\circ$ zenith angle.}
    \label{spectra}
\end{figure}

Fig. \ref{spectra} shows the primary and secondary spectrum models of GRB 190114C, for $B=10^{-17}\;\mathrm{G}$ and $\lambda_B=1\;\mathrm{Mpc}$ in $4$ time intervals, as well as the CTAO sensitivity for $30$ minutes of observation in the North and 5 hours in the South. For each time interval, the primary spectrum is shown as a dashed line while the total spectrum, which accounts for the secondary photons, is shown in solid lines. The spectra are initially dominated by the power-law primary emission with a clear EBL absorption around $200\;\mathrm{GeV}$. The flux decreases over time, and despite absorption, the cascade contribution becomes visible after an hour and still dominates after one day (red curves). For the chosen magnetic strength, the secondary emission is expected to peak at around $100\;\mathrm{GeV}$. The orange data points between $62$ and $2454\mathrm{s}$ illustrate why MAGIC could not constrain the secondary excess. It becomes potentially accessible to CTAO in the next $1800$ seconds, as indicated by the CTAO-North sensitivity curve for an hour of observation. The CTAO-South sensitivity curve for $5$ hours is also below the expected cascade of the following night (red line), indicating that the IGMF signature could stay visible the second night. We verified, using the same methodology, that CTAO remains sensitive to IGMF down to $10^{-19}\;\mathrm{G}$.

\begin{figure*}
    \centering
    \includegraphics[width=1\linewidth]{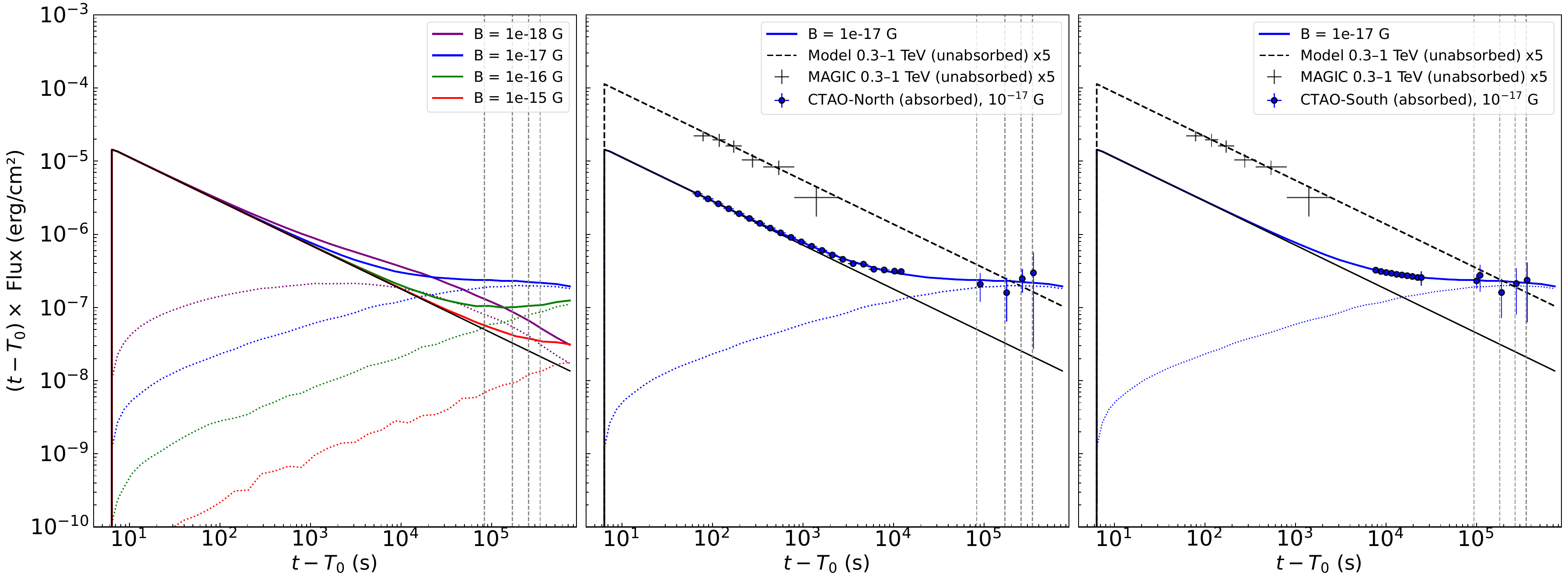}
    \caption{\textbf{(a)} Light curves of GRB 190114C multiplied by the elapsed time, showing the primary emission in the CTAO energy band ($110\,\mathrm{GeV} - 10\,\mathrm{TeV}$; solid black line) and the total emission (primaries plus cascade) for various IGMF strengths (coloured lines), and cascade contributions (dotted lines). The vertical dotted lines indicate the $2^\mathrm{nd}$ to $5^\mathrm{th}$ nightfalls (the burst occurred during the first night). \textbf{(b)} Total emission for $B = 10^{-17}\,\mathrm{G}$ (blue lines) compared to CTAO-North simulated data points (blue circles) in the analysis time bins. \textbf{(c)} Same as (b) but for CTAO-South. For illustration purposes, the MAGIC data points (black crosses) and the intrinsic model in the MAGIC energy band (dashed line) are shown in (b) and (c), offset by a multiplicative factor of five for clarity.}
    \label{lightcurve}
\end{figure*}

Fig. \ref{lightcurve} (a) shows the light curves of the primary and total fluxes of the source in the $110\;\mathrm{GeV}-10\;\mathrm{TeV}$ band for different $B$ values. These curves are multiplied by the elapsed time to better highlight the late times where cascade starts to dominate the flux. Fig. \ref{lightcurve} (b) and Fig. \ref{lightcurve} (c) show the reconstructed flux (times the elapsed time) in the North and South, respectively. The North observation is one hour longer than in the South. From the third night, the total flux falls below the CTAO sensitivity at $100\;\mathrm{GeV}$ for five hours of observation ($2\times 10^{-12}\;\mathrm{erg}\cdot \mathrm{s}^{-1}\mathrm{cm}^{-2}$), explaining the larger error bars. After $T_0+1\mathrm{h}$, the secondary fluxes can dominate, starting with the cascade generated by the weakest magnetic fields that induce the most intense cascades, although with a faster decay. Indeed, the peak of the cascade shifts rapidly towards energies below the CTAO energy range. For $B=10^{-17}\;\mathrm{G}$, a visible secondary flux is expected to be observable from the first night on, and for a whole week.  For $B=10^{-16}\;\mathrm{G}$, the secondary flux excess appears only from the second night. Therefore, these qualitative results tend to indicate that CTAO is sensitive to IGMF strengths as strong as $10^{-16}\;\mathrm{G}$.

\subsubsection{Spectral-Temporal fit: reference study}
\label{spectral-temporal}
Moving beyond qualitative expectations, we perform a spectral-temporal analysis to quantify the precision with which CTAO can reconstruct IGMF and GRB parameters. In this section, we simulate the CTAO full array response to GRB 190114C for several $B$ values in semi-realistic conditions. In the next two sections, we assess in the robustness of these constraints under varying assumptions and observational conditions. This systematic approach allows us to define robust boundaries of the CTAO sensitivity to the IGMF.

\begin{figure*}
    \centering
    \includegraphics[width=1\linewidth]{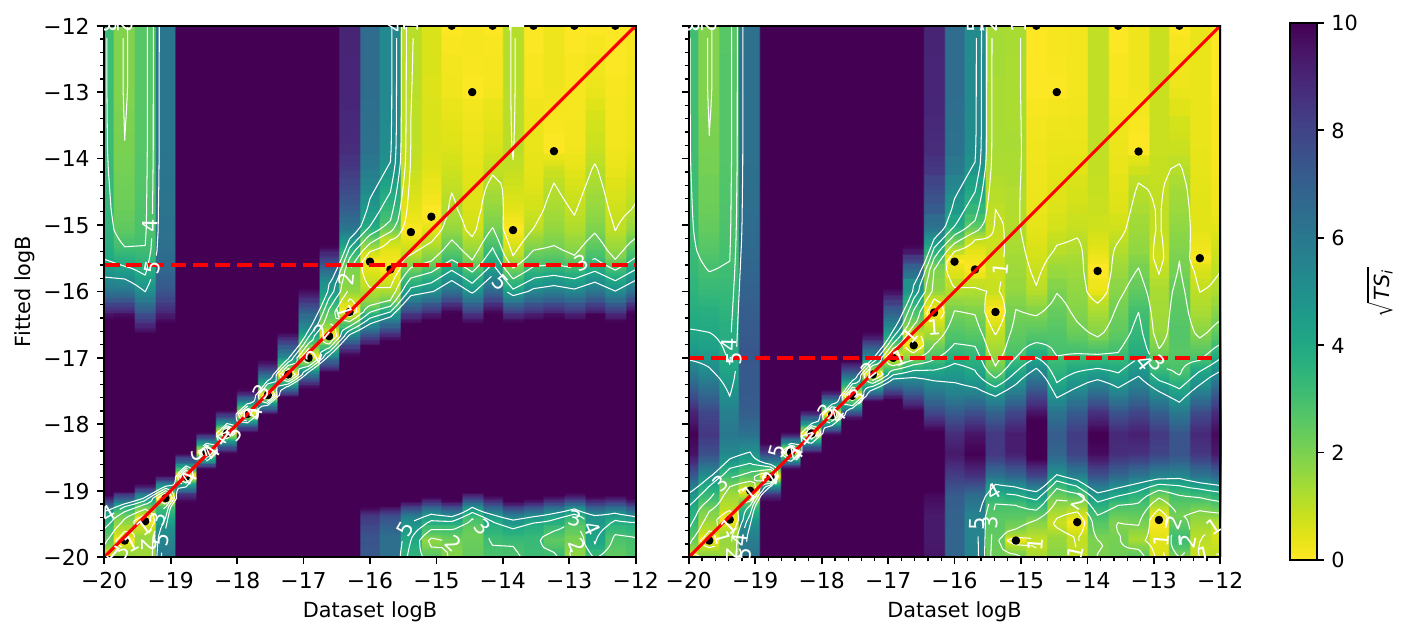}
    \caption{Likelihood maps of GRB 190114C at $\lambda_B = 1\;\mathrm{Mpc}$, with simulated $B$ values on the abscissas and fitted values on the ordinates. Best-fit values are shown as black dots. The colour map represents uncertainty levels up to $10\sigma$, with contours in white. The dashed red line highlights the lower limit at about $3\sigma$ for datasets with the strongest magnetic fields. Right: $E_\mathrm{cut}$ free to be fitted down to arbitrary low energies. Left: $E_\mathrm{cut}$ free to be fitted down to at least $10\;\mathrm{TeV}$.}
    \label{FitMap_GRB19}
\end{figure*}

The studies above show that CTAO should be able to probe magnetic field strengths up to $10^{-16}\;\mathrm{G}$. To obtain more robust and precise constraints, we simulate the CTAO response over five consecutive nights in both sites. In the North, the first night is divided into $22$ logarithmically spaced time slices, while each of the following nights is considered as a single observation period. In the South, the first night is split into $10$ bins, the second into two, and the remaining three nights are treated as one slice each. The IRF is selected according to the altitude at the start of each observation. The larger time bins from the second night onward are justified by the decreasing flux and the need for sufficient statistics. Tests with smaller bin sizes confirm that the results remain stable, as the differences in the best IRF from bin to bin are smaller than the statistical uncertainties.

We fix the correlation length to $\lambda_B = 1\;\mathrm{Mpc}$ and, in the fit, we assume that the intrinsic spectrum of the GRB has a cut-off energy, $E_\mathrm{cut}$, larger than $10\;\mathrm{TeV}$, as used by other authors (\citealp{Vovk_2023a}, \citealp{Xia_2024}, and \citealp{Miceli_2024}). The simulated spectra with different magnetic field strengths are compared to the table model, with all parameters free to vary within the bounds of Table~\ref{dimensions}.  The results are shown in Fig. \ref{FitMap_GRB19} (a). For each data set with an injected $\log B$ (abscissa), the analysis returns a best-fit $\log B$ (ordinate) denoted by a black dot, derived from matching the count distributions between simulated data and models (see Fig.~\ref{on-counts}). The uncertainties are determined from the likelihood profiles (see a few examples in Fig.~\ref{profiles}). 

The fit procedure results in different constraints depending on the assumed IGMF. For intermediate field strengths ($8\times 10^{-20}\;\mathrm{G} < B <  2\times10^{-16}\;\mathrm{G}$), the secondary emission contributes significantly to the overall emission. CTAO observations can recover the assumed strengths with small statistical uncertainties. As an example, for datasets with $B = 10^{-17.5}\;\mathrm{G}$, the fit returns $\log B = -17.51 \pm 0.05$ ($1\sigma$), with $\gamma = 2.20 \pm 0.02$, $\alpha = 1.60 \pm 0.01$ and $E_\mathrm{cut}=12.1 \pm 0.9\;\mathrm{TeV}$. Outside this range of magnetic strengths, the cascade does not not contribute significantly to the overall emission. For datasets created assuming weaker fields, the peak of secondary emission moves too quickly below the CTAO energy threshold of $60\;\mathrm{GeV}$. For datasets created with stronger IGMF ($B>10^{-16}\;\mathrm{G}$), the secondary emission is so diluted that it does not contribute to the total emission either. In both cases, the absence of detectable emission can be modelled either by very weak or by very strong fields, as seen in Fig. \ref{FitMap_GRB19}. Because previous studies strongly favour larger values of the IGMF, our analysis focuses exclusively on the lower limit estimated from the $3\sigma$ contours in the strong-field regime. This limit is highlighted by the red dashed line in Fig. \ref{FitMap_GRB19}.

\subsubsection{Departure from the reference}
\label{GRB190114C_departure}

\begin{table*}
    \centering
    \begin{tabular}{l l}
        \hline\hline
        \textbf{Study} & \textbf{Lower Limit ($\mathrm{G}$)}\\
        \hline
        Reference & $2 \times 10^{-16}$ \\
        Reduced fluence: $\phi_0/5$& $2 \times 10^{-16}$\\
        \textbf{Reduced fluence:  $\phi_0/10$}& $10^{-17}$ \\
        \textbf{Reduced fluence:  $< \phi_0/10$}& no constraint\\
        $\alpha$ and $\gamma$ varying within published uncertainties& $2 \times 10^{-16}$\\
        \textbf{$\phi_0/5$ and strong variations in $\alpha$ and $\gamma$ (see text)}& $3 \times 10^{-16}$\\
        \textbf{$E_\mathrm{cut}$ let free in the fit}& $10^{-17}$ \\
        Detection delay increased up to $1\,\mathrm{hr}$ & $2 \times 10^{-16}$\\
        \textbf{Afterglow signal starts $1'$ later}& $4 \times 10^{-17}$ \\
        High zenith angle (lower energy thresholds) & $2 \times 10^{-16}$\\ 
        Moonlight data taking (increased energy thresholds) & $2 \times 10^{-16}$\\
        \textbf{Adding LSTs in the South}& $5 \times 10^{-16}$ \\
        Exchanging North and South arrays & $2 \times 10^{-16}$\\
        Change in the fit model & $2 \times 10^{-16}$\\
        Omitting South site& $2 \times 10^{-16}$\\
        \textbf{Omitting North site}& $10^{-17}$ \\
    \end{tabular}
    \caption{Varying intrinsic properties and observation strategies for GRB 190114C assuming $\lambda_B = 1\;\mathrm{Mpc}$. Bold cases indicate variations that impact the lower limit.}
    \label{tab:departure}
\end{table*}

Building on this reference case, we explore variations in the intrinsic emission properties and observational conditions to evaluate their effect on the IGMF constraints. A summary is provided in Table \ref{tab:departure}.

\begin{itemize}
\item The lower limit remains unchanged up to a flux normalisation factor five times less than the original $\Phi_0$. Below that, it decreases to $10^{-17}\;\mathrm{G}$ for a factor of $10$, and is not defined for weaker intrinsic emission.
    \item Varying $\alpha$ and $\gamma$ within their respective published uncertainties does not alter this conclusion. 
    \item  To study a more realistic case, we considered a fictitious source with a flux reduced by a factor of five, and three different values of $\alpha$ $(1.19,\; 1.37,\; 1.54)$ and $\gamma$ $(2.10,\; 2.48,\; 2.83)$. For any fixed value of $\gamma$, increasing $\alpha$ from $1.19$ to $1.54$ improves the lower limits on the IGMF. Indeed, the cascade emission dominates sooner in CTAO for large $\alpha$ while slowly decaying GRBs maintain a longer-lasting primary emission, which delays the emergence of the secondary component and weakens the IGMF constraints. At the intermediate spectral index $\gamma=2.48$, limits go from $3\times10^{-19}\,\mathrm{G}$  to $3\times10^{-16}\,\mathrm{G}$. For any fixed $\alpha$, harder spectra (lower $\gamma$) produce more stringent limits on the magnetic field, as they contain a larger VHE flux capable of initiating cascades. For the intermediate time index $\alpha=1.37$, the limit decreases from $\sim3\times10^{-16}\,\mathrm{G}$  to $\sim3\times10^{-19}\,\mathrm{G}$ when $\gamma$ varies from $2.10$ to  $2.83$.
\item The results above are robust against further increases of the  $E_\mathrm{cut}$ boundary in the fit. For instance, setting the fit minimal value at  $E_{\mathrm{cut}}>30\;\mathrm{TeV}$ has no significant impact on the fit results. Since GRB190114C has a rather soft spectrum, there are not enough photons above $10\;\mathrm{TeV}$ to change the amplitude of the secondary emission. On the contrary, when $E_\mathrm{cut}$ is left free to explore lower values of the cut-off energy, $B$ and $E_{\mathrm{cut}}$ become partly degenerate (see Sec. \ref{summary}) and the lower limits on the IGMF are degraded down to $10^{-17}\;\mathrm{G}$, as shown on the right panel of Fig. \ref{FitMap_GRB19}.
\item We also investigated the impact of starting observations later after the burst trigger. The canonical lower limits of $2\times10^{-16}\;\mathrm{G}$ was obtained with CTAO observations starting $1\;\mathrm{min}$ after the start of the afterglow, as allowed by the trigger time and the source location at that time. The constraints on the IGMF remain unchanged as long as observations a) cover time windows dominated separately by the primary or secondary emission, and b) allow for a reliable estimation of the total energy absorbed by the EBL. Late observations can be dominated by cascade emission at all times and may not meet the first condition. For GRB 190114C, the secondary emission starts contributing after about $1\mathrm{h}$, and we consistently observe that the constraints on the IGMF are only degraded if observations miss more than the first hour of the afterglow. The second condition cannot be satisfied by late observations. However, any rough observation of the early afterglow with another high-energy instrument would be sufficient to maintain the above-mentioned constraints. Here, for instance, we have assumed, like in \cite{ravasio_2019}, that the afterglow decay starts with the \textit{Fermi}-LAT detection ($6\mathrm{s}$). Assuming instead that it begins with the MAGIC observation ($62\mathrm{s}$), the fluence is $4.5$ times weaker and so is the secondary emission. In this case, the lower limits fall to  $4\times 10^{-17}\;\mathrm{G}$.
\item This GRB was observed under Moonlight, a fact that we have omitted so far. The study (\citealp{Ahnen_2017}) made with MAGIC data shows that the reconstructed energy threshold evolves like $(\mathrm{NSB}/\mathrm{NSB_{dark}})^{0.4}$ where $\mathrm{NSB_{dark}}$ is the night sky background in optimal conditions at the site, and $\mathrm{NSB}$ the value under Moonlight. During the observation of GRB 190114C, the Moon was at a $77\%$ phase and at an angular distance of $50^\circ$ from the source, corresponding to a $\mathrm{NSB}/\mathrm{NSB_{dark}}$ ratio of $5$. Assuming that the relationship obtained in \citealp{Ahnen_2017} is applicable to CTAO, this corresponds to a threshold increase from $110\;\mathrm{GeV}$ to  $210\;\mathrm{GeV}$ in the North and $60\;\mathrm{GeV}$ to $114\;\mathrm{GeV}$ in the South. Results stay unchanged under these conditions. As will be discussed later, the secondary emission goes more quickly below the CTAO energy range for a higher threshold. However, this mostly affects situations with the lowest magnetic fields, and we find that the derived lower limits are unchanged for datasets created assuming $B>10^{-16}\;\mathrm{G}$. 
\item We also fitted the data produced from our reference exponential power law spectral model (Eq. \ref{model}) with either a sharp cut-off power law,
\begin{equation}
    \Phi(E < E_\mathrm{max}, t > t_\mathrm{min}) = \Phi_0\,E^{-\gamma}\,t^{-\alpha},
\end{equation}
or a log-parabola, 
\begin{equation}
    \Phi(E, t > t_\mathrm{min}) = \Phi_0\,E^{-\gamma - q\log(E/E_0)}\,t^{-\alpha}.
\end{equation}
In comparison, the exponential cut-off model introduces a gradual suppression of the high-energy flux, allowing some emission to persist beyond $E_\mathrm{cut}$, contrary to the sharp cut-off model. Apart from slightly larger error bars, the limits are unchanged. This is because the exponential term has only a marginal impact compared to a pure power law, given the dominant EBL attenuation and the inherent degeneracy with $E_\mathrm{cut}$. As far as the log-parabola model is concerned, the curvature parameter $q$ has to be very small to match the data, making it very close to a simple power law.
\item Results are unchanged ($2\times 10^{-16}\;\mathrm{G}$) if we consider that the GRB explodes at high zenith angle in the North ($20^\circ$), or stays fixed at this altitude, which corresponds to thresholds as low as $30\;\mathrm{GeV}$. This can be understood from the fact that the secondary excess essentially peaks at $100\;\mathrm{GeV}$ (see Fig. \ref{spectra}). Changing the trigger time so that the GRB explodes during night in the South site also gives similar results. Noteworthy, adding LSTs in the South (Omega configuration) improves the lower limits up to $5\times10^{-16}\;\mathrm{G}$ regardless of the conditions.
\item Exchanging the North and South arrays does not change the IGMF lower limits. Indeed, it has similar results as increasing energy thresholds with Moonlight. If we add Moonlight on top of the array exchange, the energy threshold in the South is around $600\;\mathrm{GeV}$ where the flux is highly suppressed by the EBL absorption. Consequently, lower limits are degraded to $2\times 10^{-17}\;\mathrm{G}$. In comparison, Moonlight alone did not change the lower limits with the normal Alpha configuration, thanks to the LSTs.
\item Finally, we repeated the simulations using only the North and South CTAO sites separately. As expected, the North site, where the observation starts earlier, provides performance close to that of the full array, albeit with larger uncertainties: error bars are $10$ times larger for datasets where the magnetic field can be measured, whereas the lower limits found for datasets with larger fields are slightly weaker, at $10^{-16}\;\mathrm{G}$. As for the South site alone, it performs worse due to the longer delay before it can observe the source, limiting the primary flux exposure. Lower limits only reach $10^{-17}\;\mathrm{G}$. Combining both sites hence allows better limits for strong $B\gtrsim 10^{-17}\;\mathrm{G}$. 
\end{itemize}

\subsubsection{Variation with the correlation length}
\label{GRB190114C_lambda_B}
As discussed in Section \ref{Theory}, magnetic deflections occur in a diffusive regime when the correlation length $\lambda_B$ is much smaller than the electron cooling distance $D_\mathrm{ic}$. Because the pairs travel multiple randomly oriented magnetic zones, the deflection angle grows as a random walk, resulting in a smaller global deflection $\delta$ compared to a uniform field of the same strength. Equating the deflections in the two regimes yields a mapping $B_\mathrm{uniform} \approx B_\mathrm{diffusive} \sqrt{\lambda_B / D_\mathrm{ic}}$. Consequently, $\log B$ and $\lambda_B$ are degenerate: a strong $B$ with a small $\lambda_B$ can mimic a weaker $B$ with a large $\lambda_B$. Because $D_\mathrm{ic}$ is inversely proportional to the electron energy, this exact mapping is slightly dependent on the observation energy range (see \citealp{Neronov_2009}). At energies probed by CTAO, the transition is around $10\;\mathrm{kpc}$. A field of $\lambda_B = 1\;\mathrm{pc}$ with $B = 10^{-16}\;\mathrm{G}$ is equivalent, at first order, to $\lambda_B = 1\;\mathrm{Mpc}$ with $B = 10^{-19}\;\mathrm{G}$. 

\begin{figure}
    \centering
    \includegraphics[width=1\linewidth]{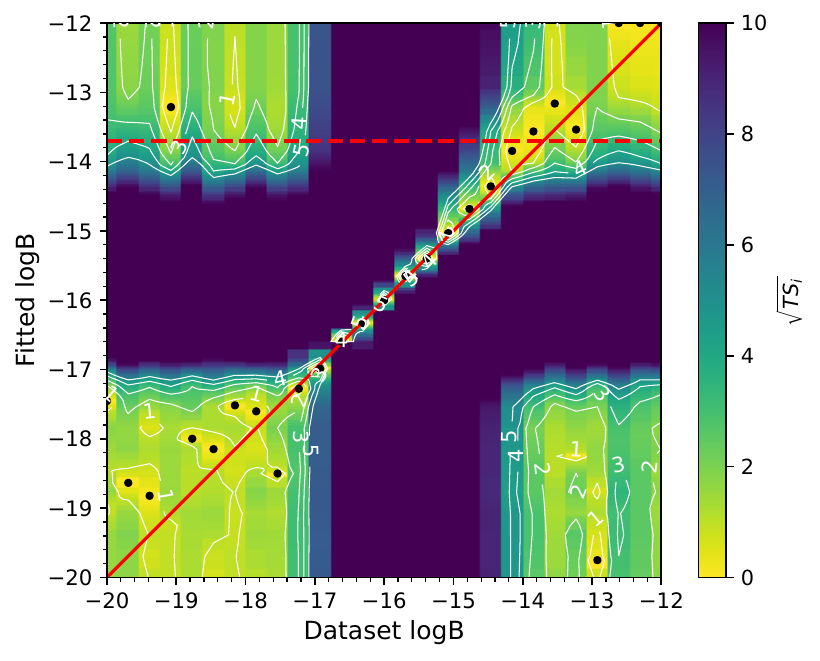}
    \caption{
    Likelihood map of GRB 190114C for $\lambda_B = 1\;\mathrm{pc}$, assuming $E_\mathrm{cut} \geq 10\;\mathrm{TeV}$. }
    \label{parsec}
\end{figure}

As a result, fitting $\lambda_B$ and $B$ simultaneously with only CTAO data remains a challenge. Instead, we repeated the full analysis for fixed $\lambda_B = 1\;\mathrm{kpc}$ and $\lambda_B = 1\;\mathrm{pc}$. Because the secondary flux grows with decreasing $\lambda_B$, we extended the observation to the sixth and eighth nights for the $1\;\mathrm{kpc}$ and $1\;\mathrm{pc}$ cases, respectively. Signal is still expected at these times, assuming such correlation lengths. As shown in Fig.~\ref{parsec}, CTAO can reach magnetic fields up to about $4\times10^{-16}\;\mathrm{G}$ for $\lambda_B = 1\;\mathrm{kpc}$ and $2\times10^{-14}\;\mathrm{G}$ for $\lambda_B = 1\;\mathrm{pc}$.  At large $\lambda_B$, CTAO is most sensitive to the weakest fields; conversely, a smaller $\lambda_B$ implies a stronger inferred $B$ for the same cascade. Hence, the lower limits obtained for $\lambda_B = 1\;\mathrm{Mpc}$ correspond to the most robust ones. 

\subsubsection{Conclusion on GRB 190114C}
\label{conclu_190114C}
Under the same conditions as MAGIC, ignoring the effect of Moonlight and assuming $E_\mathrm{cut}\geq 10\;\mathrm{TeV}$, a GRB similar to GRB 190114C would allow CTAO to measure magnetic fields up to more than $10^{-16}\;\mathrm{G}$ for $\lambda_B = 1\;\mathrm{Mpc}$ and up to $10^{-14}\;\mathrm{G}$ for smaller correlation lengths. These lower limits stay robust under an increasing of the detection threshold to account for Moonlight. This robustness partially comes from the LSTs; under Moonlight, if we exchange arrays such that the South configuration is in the North, limits drop to $2\times 10^{-17}\;\mathrm{G}$. On the other hand, lower limits improve up to $5\times 10^{-16}\;\mathrm{G}$ if we add LSTs in the South site and make the GRB burst occur during night in the South site. Alternatively, relaxing the assumption on $E_\mathrm{cut}$ degrades the lower limits to $10^{-17}\;\mathrm{G}$ for $\lambda_B=1\;\mathrm{Mpc}$ and $10^{-15}\;\mathrm{G}$ for lower $\lambda_B$. If we assume a conservative start of the afterglow ($62$ instead of $6\;\mathrm{s}$), the lower limits reduce to $4\times 10^{-17}\;\mathrm{G}$. Moreover, including realistic observation delays, for instance, up to $10\;\mathrm{min}$ after the afterglow onset, does not significantly alter the reconstructed values. Although GRBs as bright as GRB 190114C are rare, our analysis shows that the detection of a similar GRB by CTAO will allow us to establish competitive constraints on the IGMF across a broad range of conditions.

\subsection{GRB 221009A}

This GRB was detected on October $9^\mathrm{th}$, 2022, at 13:16 by Fermi GBM (\citealp{Lesage_2023}), at $\mathrm{RA}=288.28^\circ$, $\mathrm{DEC}=19.77^\circ$, with an isotropic energy estimated to $10^{55}\;\mathrm{erg}$, making it nicknamed the ``Brightest Of All Times'' (BOAT). Its redshift was measured at $z=0.151$ (\citealp{Castro-Tirado_2022}). LHAASO detected gamma rays above $300\;\mathrm{GeV}$ from $T_0+225\;\mathrm{s}$, for $6000\;\mathrm{s}$ (\citealp{Lhaaso_2023}). Because of the full Moon, H.E.S.S.\ could only derive upper limits (\citealp{Aharonian_2023_2}) with data acquired $56\;\mathrm{h}$ after $T_0$ until the $9^\mathrm{th}$ night. For the same reason, the first CTAO LST, LST-1, started taking data only $1.33\;\mathrm{days}$ after $T_0$. After the Moonlight decreased, they obtained flux upper limits for data taken between $6$ and $19$ days after $T_0$. In a later publication, after refining their analysis and getting customised IRFs taking into account the presence of the Moon,  a $4.1\sigma$ detection could be obtained during the $1.75\;\mathrm{h}$  second night, at $T_0+1.33\;\mathrm{d}$ as shown on Fig. \ref{LST1} (black points, extracted from \citealp{lst1}).

Following the LHAASO results, the GRB flux is modelled as a power-law in energy with a time evolution made of four consecutive power-laws:
\begin{equation} 
\label{GRB221009A_model} 
\Phi(E, t_i < t \leq t_{i+1}) = \Phi_i \left(\frac{E}{1 \; \mathrm{TeV}}\right)^{-2.32} \left( \frac{t}{t_{i}}\right)^{-\alpha_i}\;\exp \;(-E/E_\mathrm{cut}) 
\end{equation}
where the $\Phi_i$ are the normalisation factors in each interval are recursively defined as $\Phi_{i+1}=(t_{i+1}/t_i)^{-\alpha_i}\times\Phi_{i}$ and where the global normalisation is set so that $\frac{1}{5-0.3}\int_{0.3}^{5}\Phi(E,t)\;\mathrm{d}E = 1.31\times 10^{-5}\;\mathrm{erg}^{-1}\;\mathrm{s}^{-1}\;\mathrm{cm}^{-2}$ at $t = 18.80 \ \mathrm{s}$. We again choose $E_\mathrm{cut}=10\;\mathrm{TeV}$. Noteworthy, this exponential cut-off remains consistent with the detection of a small flux at $13\;\mathrm{TeV}$, consistent with the observations (see \citealp{cao_2023}). We still checked that assuming $E_\mathrm{cut}=13\;\mathrm{TeV}$ does not change our results. The intervals are defined by $t_i = [0,\;4.85,\;15.37,\;22,\;670,+\infty]\;\mathrm{s}$ and the respective temporal indices $\alpha_i=[14.9, 1.30,-0.05,-1.06,-2.21]$.

\subsubsection{Qualitative and quantitative analysis}
\label{BOAT}
\begin{figure*} 
\centering 
\includegraphics[width=1\linewidth]{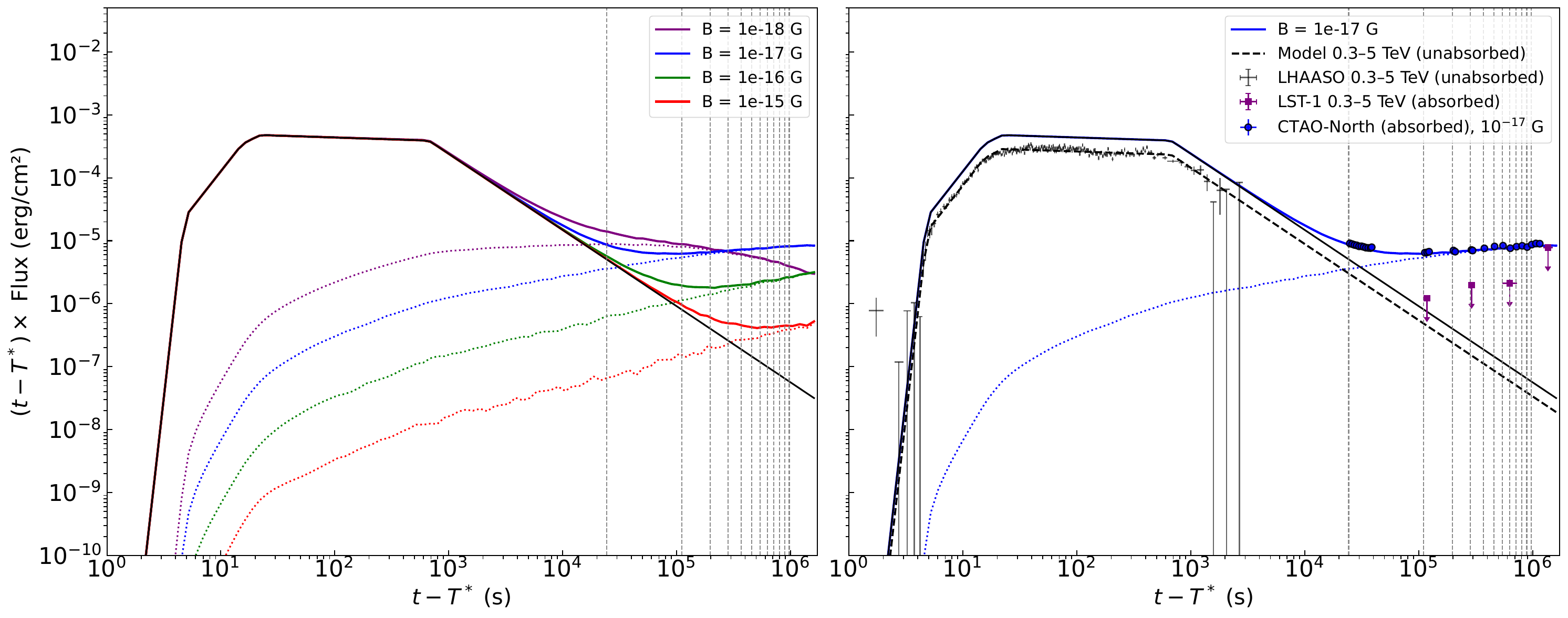} 
\caption{\textbf{(a)} Light curve of GRB 221009A between $40\;\mathrm{GeV}$ and $10\;\mathrm{TeV}$, showing the primary emission (solid black line) and total emission (primaries plus cascade) for various IGMF strengths (coloured lines), starting at $T^*\equiv T_0+225\mathrm{s}$. \textbf{(b)} Total emission for $B = 10^{-17}\,\mathrm{G}$ (blue line) compared to CTAO-North data points (blue circles) integrated over the $40\;\mathrm{GeV}-10\;\mathrm{TeV}$ energy band. The LHAASO points (black dots; extracted from \citealp{Lhaaso_2023}) and model in the LHAASO band (black dashed line) are shown for illustration purposes. The LST-1 upper limits are in purple.} 
\label{lightcurve2} 
\end{figure*}

The modelled primary light curve as well as the total flux expectations for a selection of strengths in the range $10^{-18}$ to $10^{-15}\;\mathrm{G}$ are represented on the left plot of Fig. \ref{lightcurve2}, showing that the cascade is potentially observable for weeks or months at levels largely above the primary flux. The right plot shows our emission model in the range $[300\;\mathrm{GeV}, 5\; \mathrm{TeV}]$ for $10^{-17}\;\mathrm{G}$, and for illustration, the LHAASO measurements, corrected for the EBL absorption as given in the \cite{Lhaaso_2023}. The blue circles represent the total reconstructed flux between  $40\;\mathrm{GeV}$ and $10\;\mathrm{TeV}$, at the times used in our analysis. The purple data points represent the LST-1 upper limits (\citealp{lst1}). From a rough comparison to our simulated cascades, we can exclude $B$ values from $3 \times 10^{-18}$ to $3 \times 10^{-17}\;\mathrm{G}$ using LST-1 data only. 

\begin{figure} 
\centering 
\includegraphics[width=1\linewidth]{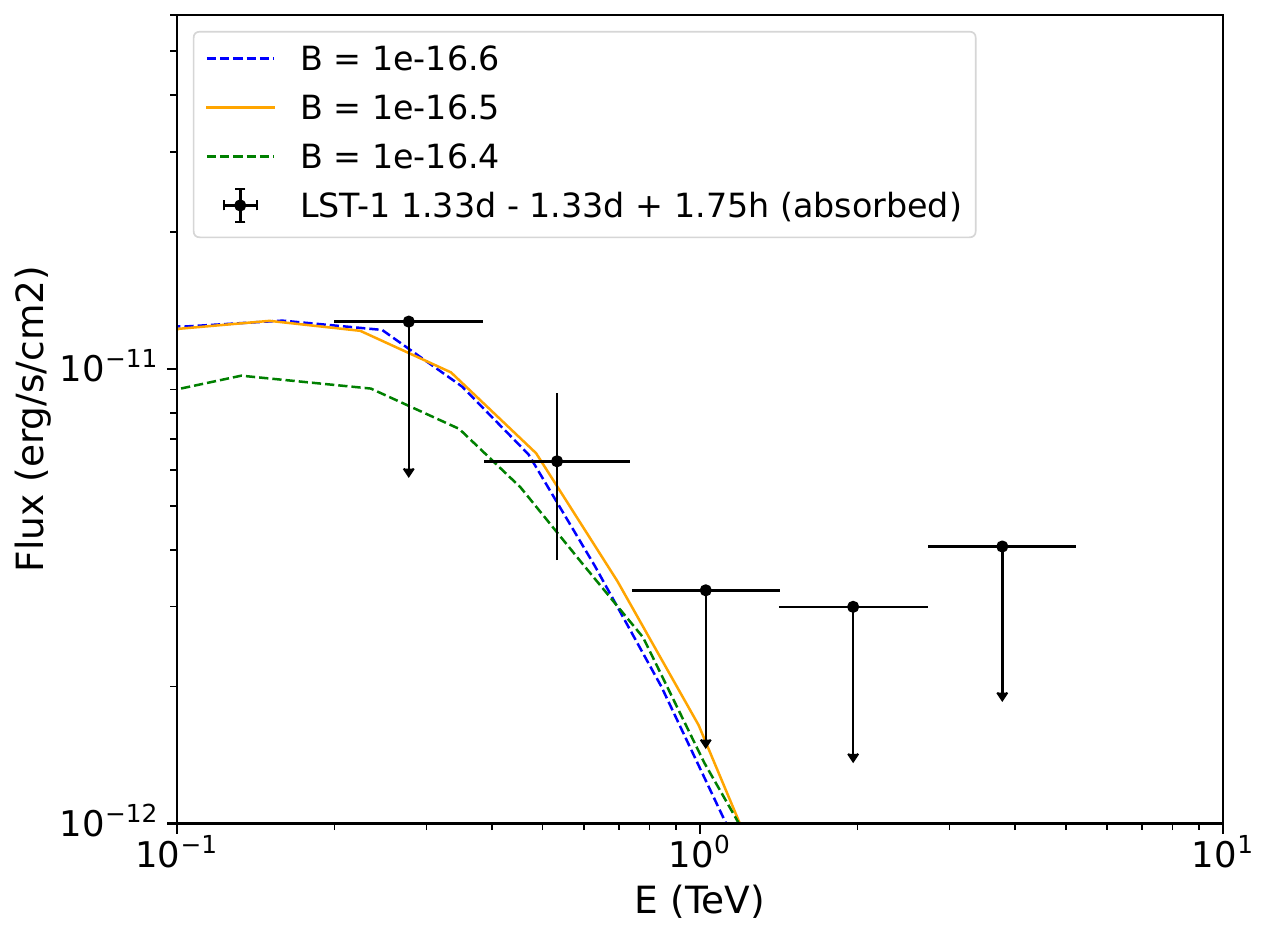} 
\caption{Published LST-1 data for the second night and simulated energy spectra for various $B$ values. The cascade for the best-matching value, $10^{-16.5}\;\mathrm{G}$, is in solid line. For illustration, two models with field strengths close to this value are drawn in dashed lines.} 
\label{LST1} 
\end{figure}

Regarding the spectra, the data shown on Fig. \ref{LST1} are compatible with a $B$ value close to $3\times 10^{-16}\;\mathrm{G}$ (for $\lambda_B=1\;\mathrm{Mpc}$). Stronger fields would not induce enough secondary emission to reach the data point flux level, and  weaker fields would generate a too strong cascade, which would violate the upper limits. Note that this value is consistent with the lower limits at $10^{-17}\;\mathrm{G}$ established with later non-detections.

In our simulations of the full array, the GRB occurs $7$ hours before the start of observation at North site, allowing the telescopes to be on target with no slewing delay. Since we ignore Moonlight, the observations last $4$ hours per night, during which the source is mostly above $40^\circ$, corresponding to an energy threshold of $40\;\mathrm{GeV}$. The GRB is detected with a maximal cumulated significance of $65\sigma$ at $T_0+11$ hours. This rather low significance compared to GRB 190114C can be explained by the fact that CTAO would start observing $7$ hours after the burst. In the South array, the detection is possible $11$ hours after the burst and is visible for $2$ hours per night. The source evolves at lower altitudes, giving an energy threshold of $350\;\mathrm{GeV}$. The GRB is detected with a cumulated significance of $64\sigma$ at $T_0+12$ hours. Once again, the low significance can be attributed to the large delay before starting the observation.

To simulate the performance of the full CTAO array, we simulate data for $12$ days in the North (adding the South does not provide any significant gain). After $12$ days, the total flux becomes undetectable in the North array, for any IGMF value (see Fig. \ref{lightcurve2}). The first hours of the emission are missed, and only the last time index is fitted, assuming that there is no secondary signal before $670\mathrm{s}$. The result of the fit is shown on Fig. \ref{map22} . The injected $B$  value is found with a good precision between $10^{-20}$ and $10^{-15}\;\mathrm{G}$. The access to weaker strengths than for GRB 190114C is due to the lower energy threshold, $40\;\mathrm{GeV}$ instead of $100\;\mathrm{GeV}$. For stronger strengths, the large flux of the BOAT gives higher secondary emissions, and the rapid decay of its intrinsic flux reduces the contamination by the primary photons, allowing CTAO to constrain more intense fields, up to $10^{-15}\;\mathrm{G}$.

\subsubsection{More constraining observation conditions}
\label{GRB221009A_more_constraining}
\begin{figure} 
\centering 
\includegraphics[width=1\linewidth]{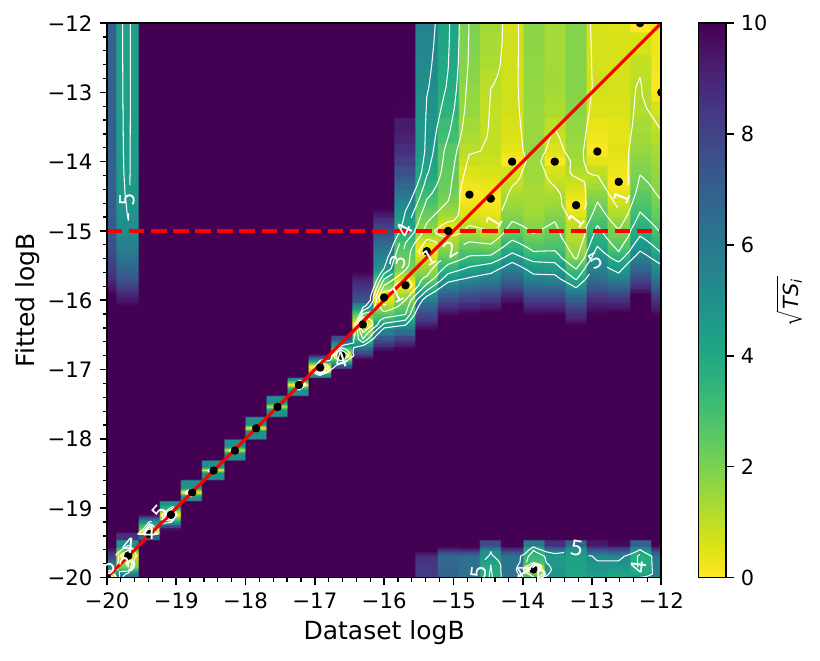} 
\caption{Likelihood map of GRB 221009A for $\lambda_B=1\;\mathrm{Mpc}$, assuming $E_\mathrm{cut}\geq10\;\mathrm{TeV}$.}
\label{map22} 
\end{figure}

If we assume strong Moonlight conditions ($100\%$ illumination, $85^\circ$ separation at most, with a night sky background $10$ times higher than the usual reference) as reported for this event, the CTAO-North energy threshold increases from $40\;\mathrm{GeV}$ ($40^\circ$ altitude, beginning of the nights) and $110\;\mathrm{GeV}$ ($60^\circ$ altitude, end of the nights) to $100\;\mathrm{GeV}$ and $\sim 300\;\mathrm{GeV}$ respectively. This corresponds to the thresholds used for the LST-1 observation, respectively at the beginning and end of the first night, reducing the lower limit to $\approx 3\times 10^{-16}\;\mathrm{G}$. We checked that fixing the energy threshold to either $100$ or $300\;\mathrm{GeV}$ does not change this lower limit. However, merely modifying energy thresholds is not a valid approximation to full-Moon conditions, as was the case for the first night of this GRB. If we instead skip the first night and apply the increased energy thresholds to the other nights, simulating observation conditions more similar to those of LST-1, fields can still be measured up to $10^{-16}\;\mathrm{G}$. In contrast to the no-Moonlight results (Fig.~\ref{map22}), however, no lower limits on $B$ can be established for stronger fields, since the degeneracy between $B$ and $E_\mathrm{cut}$ is total, even under the assumption $E_\mathrm{cut} \geq 10\,\mathrm{TeV}$.

\subsubsection{Conclusion for GRB 221009A}
\label{conclu_221009A}

\begin{table*} 
\centering 
\caption{Summary of IGMF lower limits derived from GRB 221009A in various works.} 
\label{table_221009A} 
\begin{tabular}{l lllll} 
\hline\hline 
Paper  & \thead{\cite{Vovk_2023b}} & \thead{\cite{Huang_2023}}   &\thead{\cite{Xia_2024}}&\thead{\cite{Miceli_2024}}& \thead{This work}\\[3pt] 
\hline 
$B_\mathrm{min}$ [G]  & $10^{-19}$ & $10^{-18}$   &$10^{-17}$&$10^{-17}$ (expected)& $10^{-15}$ (expected)\\[4pt] 
$E_\mathrm{cut}$ [TeV]  & 10 & 15   &10 &13& 10 \\[4pt] 
Detector& \textit{Fermi}-LAT& \textit{Fermi}-LAT&\textit{Fermi}-LAT + HAWC&CTAO& CTAO\\[4pt] 
Code  & \verb|CRPropa| & \verb|ELMAG|   &\verb|ELMAG| &\verb|CRPropa|& \verb|CascadEl| \\[4pt] 
\hline 
\end{tabular} 
\end{table*}

In the absence of a late signal, several works have already placed lower limits on the IGMF strength using GRB 221009A (see Table \ref{table_221009A}). Using \textit{Fermi}-LAT data up to $10$ days and $E_\mathrm{cut}=10\;\mathrm{TeV}$, \cite{Vovk_2023b} obtain $B>10^{-19}\;\mathrm{G}$.  \cite{Huang_2023} extended the analysis to several months with $E_\mathrm{cut}=15\;\mathrm{TeV}$ and found $B>10^{-18}\;\mathrm{G}$. The most stringent limit to date comes from \cite{Xia_2024}:  combining \textit{Fermi}-LAT data up to $50\;\mathrm{days}$ with the HAWC upper limits in the first $15\;\mathrm{h}$, they found $B>10^{-17}\;\mathrm{G}$ for $E_\mathrm{cut}=10\;\mathrm{TeV}$. Regarding CTAO expected performances, \cite{Miceli_2024}, with somewhat different assumptions but a similar cut-off energy ($13\;\mathrm{TeV}$), get a lower limit than us, at $10^{-17}\;\mathrm{G}$ for up to $9\mathrm{h}$ long observations of such GRBs. 

Noteworthy, \cite{Xia_2024} attribute a \textit{Fermi}-LAT detection at $400\;\mathrm{GeV}$, $0.4$ days after the burst, to a secondary emission. This delayed photon is compatible with a magnetic strength $B\approx 4\times 10^{-17}\;\mathrm{G}$. Using LST-1 data alone, we can exclude values between $3\times 10^{-18}$ and $3\times10^{-17}\;\mathrm{G}$, consistent with this previous study. 

More generally, in ideal observing conditions (no Moonlight), the full CTAO array would substantially improve these constraints. Owing to its low energy threshold and high sensitivity at late times, CTAO could detect the cascade contribution and measure fields as intense as $10^{-15}\;\mathrm{G}$. If the true magnetic field strength was $3\times10^{-17}\;\mathrm{G}$, as suggested by both the LST-1 detection and the delayed \textit{Fermi}-LAT photon, CTAO would be able to measure it with good precision while simultaneously fitting the intrinsic GRB parameters. If the IGMF had a larger strength than the detectable range, CTAO would impose lower limits up to $10^{-15}\;\mathrm{G}$.

Under more realistic observing conditions for this event, accounting for Moonlight, the performance is reduced. The CTAO-North energy threshold increases significantly, leading to a degradation of sensitivity to secondary emission. In this case, lower limits are degraded down to $\approx 3\times10^{-16}\;\mathrm{G}$, which still improves upon the LST-1 constraint by about one order of magnitude. If, in addition, the first night is excluded, as was effectively the case for LST-1 observations, secondary emission remains detectable only for $B \lesssim 10^{-16}\;\mathrm{G}$. For stronger fields, the cascade becomes too faint to be distinguished from the background, and no lower limit can be set, due to a complete degeneracy between $\log B$ and $E_\mathrm{cut}$. 

\subsection{GRB 180720B}
\label{GRB18}
\begin{figure*}
    \centering
    \includegraphics[width=1\linewidth]{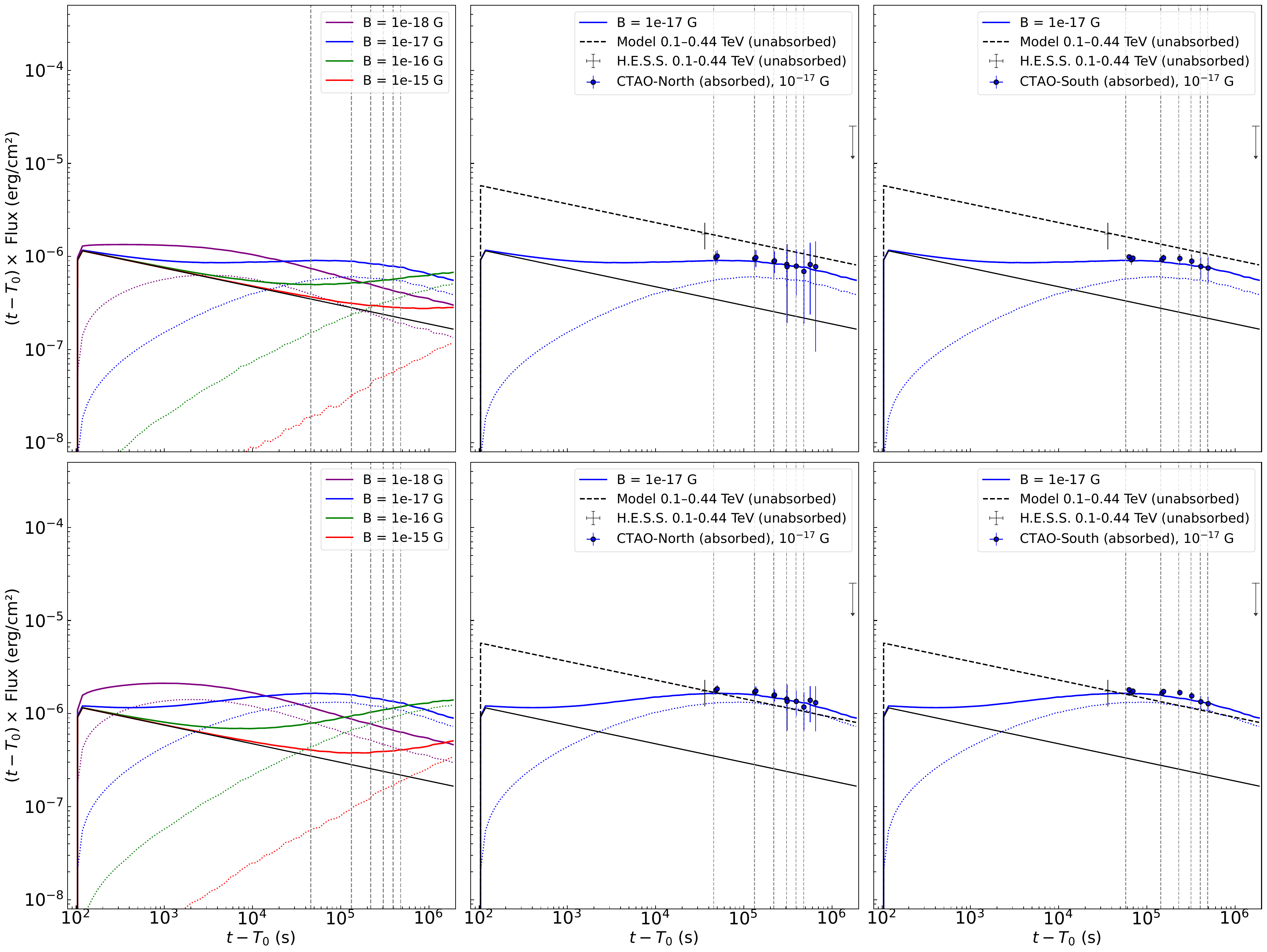}
    \caption{\textbf{Top:} \textbf{(a)} Light curve of GRB 180720B ($E_\mathrm{cut} = 10\;\mathrm{TeV}$) showing the primary emission in the CTAO energy band ($110\;\mathrm{GeV} - 10\;\mathrm{TeV}$; solid black line) and the total emission (primaries plus cascade) for various IGMF strengths (coloured lines). The corresponding cascade contributions are shown as dotted lines. \textbf{(b)} Total emission for $B = 10^{-17}\;\mathrm{G}$ (blue line) compared to CTAO-North data points (blue circles) in the analysis time bins. \textbf{(c)} Same as (b) but for CTAO-South (blue squares). For illustration purposes, the H.E.S.S. data points (black crosses) and intrinsic model (dashed line) are shown in (b) and (c). \textbf{Bottom:} Same as top for $E_\mathrm{cut} = 30\;\mathrm{TeV}$.}
    
    \label{lightcurve_18}
\end{figure*}

GRB 180720B was first observed by Fermi-GBM on July $18^\mathrm{th}$, 2018 at 14:21 (\citealp{Roberts_2018}), at $\mathrm{RA}=0.53^\circ$, $\mathrm{DEC}=-2.92^\circ$, with an isotropic energy $E_{\mathrm{iso}}=6\times 10^{53}\;\mathrm{erg}$ in $[50,300]\;\mathrm{keV}$ and a redshift of $z=0.653$ measured by the Very Large Telescope (\citealp{Vreeswijk_2018}). The H.E.S.S. telescope measured a single flux point between $100$ and $440\;\mathrm{GeV}$ during $2$ hours on the first night, with a significance of $5\sigma$ (\citealp{Abdalla_2019}). After a long Moonlight period, an upper limit could be obtained at  $T_0+18\;\mathrm{days}$. At the CTAO North site, the burst occurs $12$ hours before night, and remains visible for $1\mathrm{h}30$ per night and at low altitude. It is detected at $10\sigma$, after accumulating data up to $1.60\;\mathrm{days}$ after $T_0$. In the South, the GRB is observed for $3\mathrm{h}30$ with a delay of $16$ hours after the burst, and with a maximal significance of $23\sigma$ after $1.77\;\mathrm{days}$.

So far, every VHE GRBs has been consistent with a unique temporal decay index from X-rays to VHE gamma rays (\citealp{2019} for GRB 190114C, \citealp{Hess_2021} for GRB 190829A, \citealp{Lhaaso_2023} for GRB 221009A), and in this particular case, \textit{Fermi}-LAT and Swift data are consistent with an index equal to $1.2$ in their respective energy ranges. In the absence of VHE information, we therefore assume $\alpha=1.2$. We verified that choosing $\alpha=1$ , the mean time index for \textit{Fermi}-LAT long GRBs afterglows, would only slightly worsen the results (the larger the temporal index, the sooner the cascade dominates).

We assumed that the high-energy afterglow emission started at the same time as in the \textit{Fermi}-LAT band ($100\;\mathrm{MeV}-10\;\mathrm{GeV}$), at $t=100\mathrm{s}$. The synthetic datasets are generated according to Eq. \ref{model} with $\gamma=1.60$, $\alpha=1$ and $t_\mathrm{min}=2\;\mathrm{h}$. The normalisation factor $\Phi_0$ is chosen such that $\Phi(E=154\;\mathrm{GeV},t= 2\mathrm{h})= 7.52\times10^{-10}\;\mathrm{TeV}^{-1}\;\mathrm{s}^{-1}\;\mathrm{cm}^{-2}$, the flux measured by H.E.S.S. in the $100-440$ GeV band (\citealp{Abdalla_2019}). 

GRB 180720B is more luminous than GRB 190114C but further away, resulting in a fainter source. Its high redshift is also responsible for high absorption, leaving very little space in energy for detection, between the CTAO threshold and the EBL cut-off. Its light curve was simulated over six nights after which the flux falls below the CTAO sensitivity ($2\times 10^{-12}\;\mathrm{erg}\cdot \mathrm{s}^{-1}\mathrm{cm}^{-2}$ at $100\;\mathrm{GeV}$ for $5$ hours of observation). It is shown on Fig. \ref{lightcurve_18} (a). The large uncertainties found by the fit procedure does not allow for constraining the IGMF for any value of the strength assumed in the datasets.

Moreover, GRB 180720B has a very hard energy spectrum ($\gamma=1.60$). This makes the results very sensitive to the cut-off energy. For instance, we also produced a data set with $E_\mathrm{cut}=30\;\mathrm{TeV}$; the secondary emission is then more prominent and quickly dominates the total emission for any IGMF strength below $10^{-15}\;\mathrm{G}$. For $B=10^{-18}\;\mathrm{G}$ it happens as early as $30$ minutes onward, as shown on the bottom figures of Fig. \ref{lightcurve_18}. However, the fit procedure does not provide better constraints in this situation because it remains impossible to distinguish the properties of the primary and secondary emissions, partly due to the large uncertainty on $E_\mathrm{cut}$.  Moreover, the CTAO observation would have started quite late ($T_0+12$ hours) so there is no strength of the IGMF for which CTAO could have been able to observe the transition from the primary- to the secondary-dominated emission, which is crucial to constrain the IGMF. This happens because the GRB spectrum is quite hard and produces a strong cascade emission. Consequently, by the time CTAO is able to observe the source, the width of the cascade excess spans the whole window of accessible energies, from $20$ to $200\;\mathrm{GeV}$.

Under more favourable conditions, namely a higher altitude, allowing for a lower energy threshold, this degeneracy between primary- and secondary emissions could have been broken. In this case, the primary component could be more clearly identified and disentangled from the secondary flux, opening the possibility to constrain the IGMF even in the presence of a dominant cascade contribution. This has been tested by placing GRB~180720B at a $20^\circ$ zenith angle, effectively lowering the energy threshold to $30\;\mathrm{GeV}$. In this configuration, we can probe magnetic field strengths up to $3\times 10^{-16}$ and $5\times 10^{-16}\;\mathrm{G}$ with $E_\mathrm{cut}=10$ and $30\;\mathrm{TeV}$ respectively. Noteworthy, this source is visible at higher altitudes in the South (above $60^\circ$ at the end of the night). The presence of LSTs in the South site could then have been enough to probe the primary spectrum of the source, given the lower energy threshold of these telescopes. If $B\gtrsim 5\times 10^{-16}\;\mathrm{G}$, the cascade is not detected and no lower limits can be established. It shows that, even with access to lower energies, the assumption that $E_\mathrm{cut}\geq 10\;\mathrm{TeV}$ is not constraining enough to break the degeneracy between $\log B$ and $E_\mathrm{cut}$. 

This kind of GRB is not as rare as GRB 221009A meaning that such an event might be expected during the lifetime of CTAO.

\subsection{GRB 190829A}

\begin{figure*}
    \centering
    \includegraphics[width=1\linewidth]{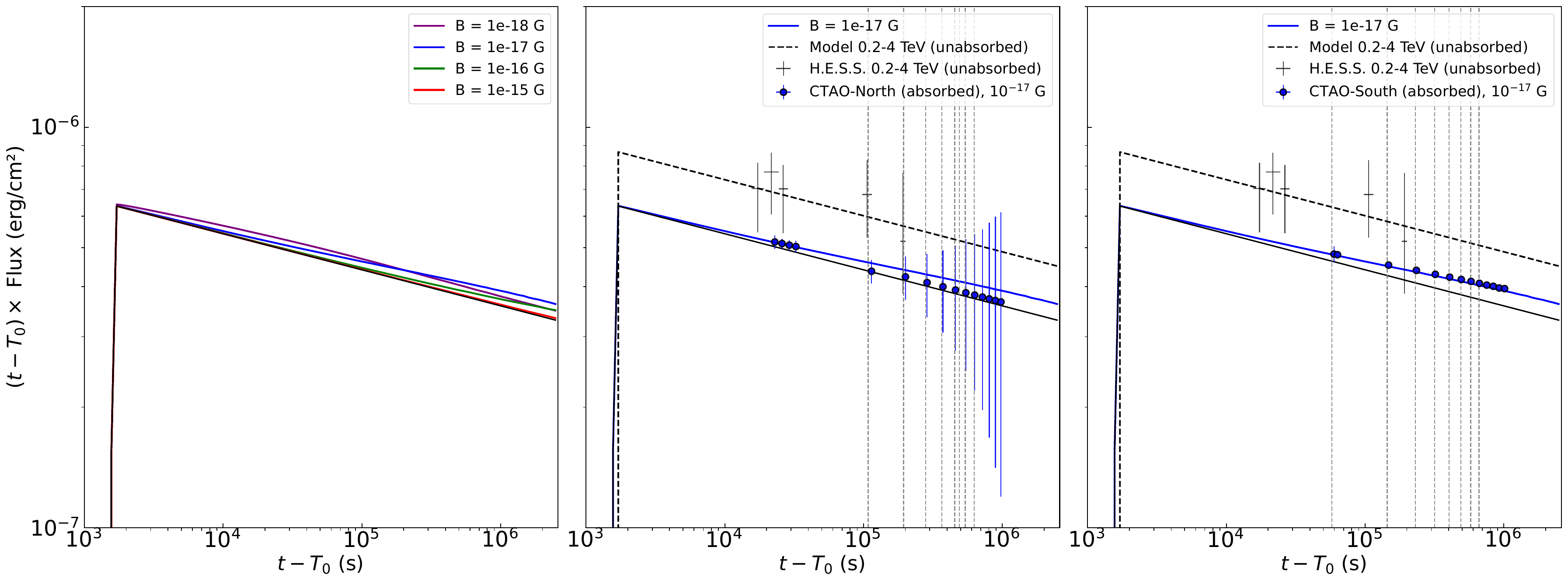}
    \caption{\textbf{(a)} Light curve of GRB 190829A showing the primary emission in the CTAO energy band ($110\,\mathrm{GeV} - 10\,\mathrm{TeV}$; solid black line) and the total emission (primaries plus cascade) for various IGMF strengths (coloured lines). The corresponding cascade contributions are shown as dotted lines. \textbf{(b)} Total emission for $B = 10^{-17}\,\mathrm{G}$ (blue line) compared to CTAO-North data points (blue circles) in the analysis time bins. \textbf{(c)} Same as (b) but for CTAO-South (blue squares). For illustration purposes, the H.E.S.S. data points (black crosses) and intrinsic model (dashed line) are shown in (b) and (c).}
    \label{lightcurve_19}
\end{figure*}

GRB 190829A was detected by Fermi-GBM on August $29^\mathrm{th}$, 2019 at 19:55 at $\mathrm{RA}=44.54^\circ$, $\mathrm{DEC}=-8.98^\circ$ (\citealp{Lesage_2019b}), with an isotropic equivalent energy of $E_\mathrm{iso} = 2\times10^{50}$ in the Fermi GBM band. The source is located at redshift $z=0.079$ (\citealp{Valeev_2019}). It was detected by H.E.S.S. between $4$ and $56\mathrm{h}$ after the burst across three consecutive nights, in the $200\;\mathrm{GeV}-4\;\mathrm{TeV}$ band (\citealp{Hess_2021}). H.E.S.S reached a mean significance of $21.7\sigma$ during the first night. Regarding CTAO, the burst occurs one hour before nightfall in the Northern site, and remains under the minimal altitude of $24^\circ$ for $4$ more hours, leaving only $3.5$ hours per night to observe the source at altitudes below $50^\circ$. In the South the burst happens $16$ hours before nightfall, and is visible $1\mathrm{h}30$ per night. It is detected up to $154\sigma$ after $9$ hours and $238\sigma$ after $1.57\;\mathrm{days}$ of accumulated data in the North and South respectively. The source is modelled as a power-law in energy and in time with $\gamma=2.07$, $\alpha=1.09$  $t_\mathrm{min}=0.44\;\mathrm{h}$. The normalisation is  $4.06\times 10^{-11}\;\mathrm{erg}\;\mathrm{s}^{-1}\;\mathrm{cm}^{-2}$ as quoted by H.E.S.S. from the measured mean flux between $0.2$ and $4\;\mathrm{GeV}$, and $1560$ and $1920\mathrm{s}$.

The afterglow emission was detected $1600\mathrm{s}$ after the burst, by Swift-XRT. We construct the light curve assuming the same starting time at high energy. As we can notice in Fig. \ref{lightcurve_19}, the flux is never dominated by the cascade regardless of the strength of the IGMF. This is the consequence of a low time index, inducing the primary flux to vanish slowly. Additionally, the VHE flux of the source is weakly absorbed due to its low redshift. As a result, no constraints on the IGMF could be obtained even with a simulation of $8$ nights. 

\subsection{Key parameters}
\label{summary}
From the limited set of detections at very high energies studied above, we can derive trends and conclusions regarding the CTAO capability to measure the IGMF strength or set limits on its existence. 
Our analysis shows CTAO can cover strengths as strong as $10^{-15} \; \mathrm{G}$ for the brightest GRBs, and  $10^{-17} \;\mathrm{G}$ for sources ten times fainter than GRB 190114C.  \\

\noindent \underline{Cut-off energy:} \\
Our results depend on the assumption of the presence and value of the intrinsic cut-off energy, which could not be measured so far because of the EBL absorption (see Section \ref{spectral-temporal}). For cut-off energies below $10\;\mathrm{TeV}$, the secondary emission peaks below the CTAO energy thresholds (see Eq. \ref{energy}), affecting the precision of our estimate. For a cut-off above $10\;\mathrm{TeV}$, a value in agreement with the recent GRB 221009A observations by LHAASO, the sensitivity to the cut-off depends on the source hardness. For soft sources, the dependence is weak, as shown for GRB 190114C ($\gamma=2.2$) and GRB 221009A ($\gamma=2.3$). In that case, the choice of $E_\mathrm{cut}$ does not affect the fit results as long as it is above $10\;\mathrm{TeV}$. On the other hand, the secondary emission from hard sources like GRB 180720B ($\gamma = 1.6$) is mostly produced by primaries at the VHE edge of the spectrum and is very sensitive to the precise cut-off. \\

\noindent \underline{Energy threshold and range:} \\
Depending on altitude, the energy threshold can be as low as $30\;\mathrm{GeV}$ in CTAO. In general, it should be below $100\;\mathrm{GeV}$ to detect the secondary emission before it becomes too low in energy (see \ref{GRB190114C_departure}). Note that the EBL cut-off reduces the effective detection energy range (at $z=0.5$, the cut-off is $150\;\mathrm{GeV}$), and, as we checked, placing our GRBs at higher redshifts, the performance degrades quickly for sources with redshift above $1$ (EBL cut-off at $100\;\mathrm{GeV}$). For high redshift GRBs, the perspective of a joint fit with Fermi at lower energies (from $10\;\mathrm{MeV}$ to $10\;\mathrm{GeV}$) is crucial to characterise the primary spectrum in the absence of the secondary gamma rays.  \\

\noindent \underline{Time, start and duration:} \\
The time evolution of the primary emission is essential: a slower decay keeps producing new secondary particles for a longer time, whereas the cascade emission starts dominating earlier with a faster decay. Since both the primary and secondary emissions become undetectable after a few days, GRBs with faster decays provide the best constraints. The observation start time has a major influence on the uncertainties for all strengths (see Section \ref{spectral-temporal}). More generally, an early characterisation of the afterglow is crucial for obtaining robust constraints. The detection of GRB 221009A by LHAASO is exemplary in this context. The observation should also last long enough, as the key point is to observe the appearance of the secondary emission on top of the intrinsic light curve. In particular, for GRBs that CTAO would have missed at the start (e.g. during the daytime), information from other observatories could be crucial. However, the rising phase of the GRB afterglow contributes negligibly to the cascade, since it does not substantially alter the total energy deposited in the high-energy band. \\

\noindent \underline{Effect of the sites:} \\
The main difference between the two CTAO sites is the presence of LSTs in the North, as their low energy threshold ($30\;\mathrm{GeV}$) is ideal for a good characterisation of the primary emission. For instance, with GRB 190114C, CTAO maintains stable lower limits at $2\times 10^{-16}\;\mathrm{G}$ even under Moonlight conditions ($210\;\mathrm{GeV}$ threshold in the North). In comparison, exchanging arrays such that the South configuration is in the North causes limits to drop to $2\times 10^{-17}\;\mathrm{G}$ under Moonlight ($600\;\mathrm{GeV}$ threshold). This disparity is further highlighted by the following experiment: if we make the GRB burst occur during night in the South, results stay unchanged in the Alpha configuration, while lower limits increase up to $5\times 10^{-16}\;\mathrm{G}$ if we add LSTs in the South site (Omega configuration).

\section{Conclusion}
In this study we present CTAO sensitivity estimates to the IGMF using afterglows of the few long VHE GRBs detected so far by current instruments. Our study encompasses, for the first time, the properties of the primary emission, the spectral and temporal evolution of the EBL-induced cascade, and several aspects related to the observation conditions as well as the modelling choices. Our work indicates that the performance of the IGMF characterisation is driven by the sensitivity to delayed secondary emission and the source brightness rather than variations in the spectral and temporal shapes of the GRB. In this respect, the ideal VHE GRB candidate should have a redshift ($0.1 <
z < 0.6$), a large temporal index so that the cascade quickly dominates over the primary spectrum, and a low spectral index to emit numerous high-energy photons. However, even with the best possible source, the potential constraints on the IGMF also depend significantly on the observation conditions.

\begin{figure}
    \centering
    \includegraphics[width=1.03\linewidth]{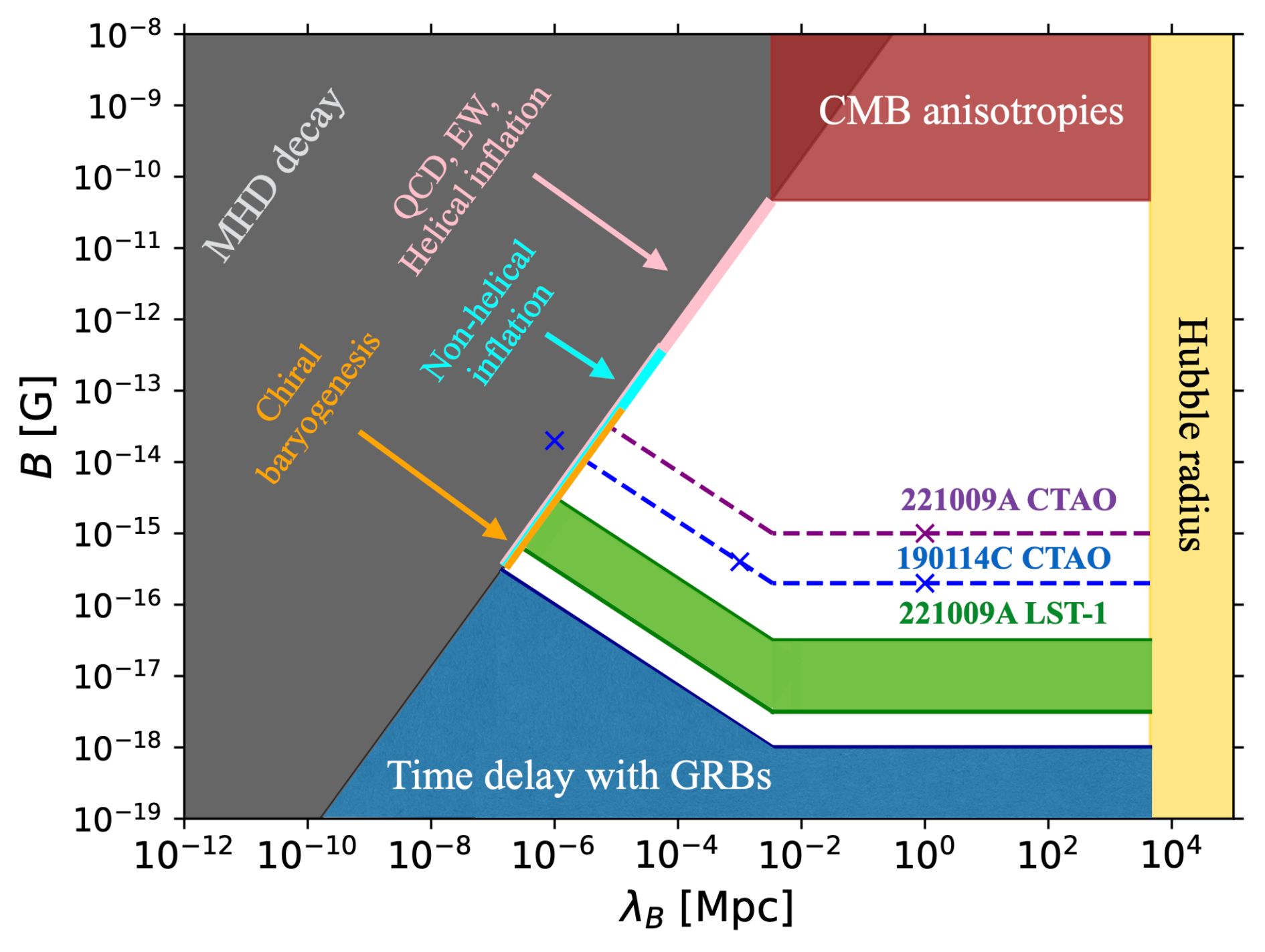}
    \caption{Theoretical and observational constraints on the IGMF. Excluded regions are coloured: early Universe theoretical predictions (dark grey); CMB anisotropies (red);  time-delay constraints from earlier works (blue); our LST-1 exclusion region (green). The expected strength lower limits depending on the type of GRB CTAO will observe are the dashed blue and purple lines, with crosses representing points for which we performed the complete analysis.}
    \label{igmf}
\end{figure}

Fig. \ref{igmf} shows theoretical predictions and observational constraints obtained so far, as well as our results, in the strength-correlation length plane.
The coloured regions represent excluded areas. In the early Universe, interactions with the original ambient plasma cause the primordial magnetic field to dissipate, primarily at small scales, leading to a strength decay. This explains the grey area, which tends to expand with time as indicated by the arrows. The red area corresponds to upper limits from CMB anisotropies (mostly coming from clumping at recombination). It represents the maximum magnetic field strength compatible with the observed CMB power spectrum. Fields exceeding this limit would have generated distortions in baryon acoustic peaks (resulting from the primordial plasma dynamics), which are not observed in current data (see \citealp{Jedamzik_2019}). The yellow-shaded area corresponds to the Hubble radius, set at $4.420\;\mathrm{Gpc}$. The current strongest time-delay constraints with GRBs are illustrated in blue (\citealp{Dzhatdoev_2023}). The break at lower correlation lengths corresponds to the transition to the "random walk" regime, which occurs when $\lambda_B$ exceeds the lepton cooling distance (see \ref{Theory}). 

The two curves show our results, the crosses indicating our simulation points with $E_\mathrm{cut}=10\;\mathrm{TeV}$. The dashed lines indicate the extrapolated constraints in the diffusive regime at smaller correlation lengths, assuming (see Sections \ref{Theory} and \ref{GRB190114C_lambda_B}). They show the upper bounds of the range accessible to CTAO for GRB 190114C (see Section \ref{spectral-temporal}) and GRB 221009A (see Sec. \ref{BOAT}), respectively. They also correspond to the lower limits CTAO would obtain in the strong field regime. For good observation conditions, we obtain lower limits one or two orders of magnitude higher than what the \textit{Fermi}-LAT energy domain can provide, as expected (see Section \ref{Theory}).
At shorter correlation lengths ($\lambda_B \gtrsim 1\;\mathrm{pc}$), the possibility for CTAO to probe magnetic fields up to $10^{-13}\;\mathrm{G}$  raises some potential for indirectly probing Early Universe phenomena such as the baryogenesis, in particular a chiral-induced baryogenesis (orange segment, see \citealp{Joyce_1997} and \citealp{Huber_2006}) and more marginally non-helical primordial fields generated during inflation (cyan segment, see \cite{neronov2020}).

The green region, excluded using published LST-1 data on GRB 221009A (see \cite{lst1}), reaches intensities up to $3\times 10^{-17}\;\mathrm{G}$ at $\lambda_B=1\;\mathrm{Mpc}$ assuming $E_\mathrm{cut}=10\;\mathrm{TeV}$. If we extrapolate the results to $\lambda_B=1\;\mathrm{pc}$, they also exclude a large fraction of the chiral-induced baryogenesis models. 
On the other hand, we have checked that the $4.1\sigma$ data point observed during the second night is compatible with a magnetic field strength $B = 3\times10^{-17}\;\mathrm{G}$ for $\lambda_B = 1\;\mathrm{Mpc}$ and $E_\mathrm{cut} = 10\;\mathrm{TeV}$. This value is consistent with all previous constraints, including the most conservative limits derived from blazar observations. It could therefore be the first indication of the presence of an IGMF field. With $\lambda_B\approx 1\;\mathrm{pc}$, the LST-1 data indicate a strength of $10^{-15}\;\mathrm{G}$, consistent with either inflationary or phase-transition scenarios. However, it requires a refined analysis and additional detections. As we show here, the full CTAO array can confirm this indication by detecting events similar to GRB 190114C, which should be far more frequent than GRB 221009A, estimated to be a millennial event.

All the statements so far are based on a strong assumption that the GRB high-energy component decreases sharply after $E_\mathrm{cut} = 10\;\mathrm{TeV}$. A higher value would induce brighter cascades, improving all lower limits and enlarging the LST-1 exclusion area. A smaller cut-off energy would have opposite consequences: limits would decrease and the LST-1 exclusion area would shrink.
In principle, IGMF studies might also be affected by EBL uncertainties. These were not taken into account here, as the best targets for IGMF studies are moderately distant GRBs (particularly the ones studied here), where EBL properties are well-constrained and most models agree. However, such uncertainties must be investigated if a higher-redshift GRB is used.
Finally, our study also does not account for the prompt component, which has never been detected in the $\mathrm{TeV}$ energy range. Since the prompt component acts as a Dirac pulse or a very fast decaying afterglow, it would likely increase the secondary flux, if bright enough, without changing the underlying primary spectrum, increasing the expected IGMF lower limits.

In general, a good characterisation of the underlying primary spectrum is crucial for distinguishing the secondary excess. CTAO suffers from the fact that data are acquired at night only, thus potentially hiding the GRB at early times or else creating gaps in the detected light curve. This can be compensated, in the Northern Hemisphere by HAWC, LHAASO and VERITAS, and in the Southern Hemisphere by H.E.S.S. and the future SWGO facilities.

Still, despite accounting for more realistic conditions, we cover a wider IGMF range than previous publications (\citealp{Miceli_2024}, \citealp{Vovk_2023a}, \citealp{Xia_2024}). Using GRBs 190114C and 221009A, and assuming a cut-off energy of $10\;\mathrm{TeV}$ and a correlation length of $1\;\mathrm{Mpc}$, we derive lower limits reaching $10^{-15}\;\mathrm{G}$  in ideal conditions, and $10^{-16}\;\mathrm{G}$ when we account for more realistic observations that induce higher detection energy thresholds. In comparison, GRBs 180720B and 190829A offer a limited opportunity to constrain the IGMF strength. An extensive study of observation conditions and GRB brightness let us conclude that CTAO is guaranteed to probe fields up to $4 \times 10^{-17}\;\mathrm{G}$ at least. Noteworthy, even in the case of a non-detection, CTAO would independently corroborate the exclusion ranges established by previous instruments, effectively confirming the exclusion of very weak fields. The crucial point for constraining or characterising an intergalactic magnetic field, the possible vestige of a primordial field, is the limited number of VHE GRBs detected, a situation that will evolve with the operation of CTAO over the coming decades. Notably, an exhaustive population-based study could refine this conclusion.

\begin{acknowledgements}
This work was conducted in the context of the CTAO consortium and has used the CTAO instrument response functions. We acknowledge the review from the CTAO Science working groups, the Speaker and Publication Office, and, in particular, Andrii Neronov, Reshmi Mukherjee and Zeljka Bosnjak.

\end{acknowledgements}

\bibliographystyle{aa}
\bibliography{refs}

\end{document}